\documentclass[prd,aps,floatfix,nofootinbib,twocolumn,tightenlines,showpacs]{revtex4-1}
\usepackage{dcolumn}% Align table columns on decimal point
\usepackage{latexsym}
\usepackage{graphicx}
\usepackage{bm}
\def\G{\Gamma}

\def\co{{\cal O}}

\def\bx{{\bf x}}

\def\svev#1{\left\langle #1\right\rangle}       % variable < >
\def\tr{{\rm tr}\,}

\long \def \blockcomment #1\endcomment{}

\def\Eq#1{Eq.~(\ref{#1})}

\newcommand{\bee}{\begin{equation}}
\newcommand{\ee}{\end{equation}}
\newcommand{\beea}{\begin{eqnarray}}
\newcommand{\eea}{\end{eqnarray}}

\begin{document}
%%%%%%%%%%%%%%%%%%%%%%%%%%%%%%%%%%%%%%%%%%%%%%%%%%%%%%%%%%%%%%%%%%%%%%

\title{
Running coupling and mass anomalous dimension of SU(3) gauge theory with two flavors of symmetric-representation fermions
}
\author{Thomas DeGrand}
%\email{degrand@pizero.colorado.edu}
\affiliation{Department of Physics,
University of Colorado, Boulder, CO 80309, USA}
\author{Yigal Shamir}
%\email{shamir@post.tau.ac.il}
\author{Benjamin Svetitsky}
%\email{bqs@julian.tau.ac.il}
\affiliation{Raymond and Beverly Sackler School of Physics and Astronomy, Tel~Aviv University, 69978 Tel~Aviv, Israel}

\begin{abstract}
We have measured the running coupling constant of SU(3) gauge theory coupled to $N_f=2$ flavors of symmetric representation fermions, using the Schr\"odinger functional scheme.
Our lattice action is defined with hypercubic smeared links which, along with the larger lattice sizes, bring us closer to the continuum limit than in our previous study.
We observe that the coupling runs more slowly than predicted by asymptotic freedom, but we are unable to observe fixed point behavior before encountering a first order transition to a strong coupling phase.
This indicates that the infrared fixed point found with the thin-link action is a lattice artifact.
The slow running of the gauge coupling permits an accurate determination of the mass anomalous dimension for this theory, which we observe to be small, $\gamma_m \alt0.6$, over the range of couplings we can reach.
We also study the bulk and finite-temperature phase transitions in the strong coupling region.
\end{abstract}

\maketitle

%%%%%%%%%%%%%%%%%%%%%%%%%%%%%%%%%%%%%%%%%%%%%%%%%%%%%%%%%%%%%%%%%%%%%
\section{Introduction}
%%%%%%%%%%%%%%%%%%%%%%%%%%%%%%%%%%%%%%%%%%%%%%%%%%%%%%%%%%%%%%%%%%%%%
Several possibilities for new physics beyond the Standard Model involve a new strongly interacting sector of gauge fields and light fermions.
In the oldest such idea, technicolor, interactions are presumed
to be asymptotically free at very short distance but strong at long distance, leading to a fermion--antifermion condensate $\svev{\bar\psi\psi}$
that replaces the Higgs vacuum expectation value~\cite{Hill:2002ap}.
A more recent proposal couples the standard model to a sector of ``unparticles,'' in which long distance dynamics is conformal
in the limit of zero fermion mass~\cite{Georgi:2007ek}.
 Determining whether either of these scenarios occurs in a candidate
theory is a nonperturbative question. In the last few years many groups have begun to attack it 
using lattice methods~\cite{reviews}.

One common tool for diagnosing the infrared structure of a gauge theory is its beta function.
In perturbation theory it has the expansion
\bee
\beta(g^2)=\frac{dg^2}{d\log q^2}=-\frac{b_1}{16\pi^2}g^4-\frac{b_2}{(16\pi^2)^2}g^6+\cdots,
\label{2loopbeta}
\ee
where, for an SU($N$) gauge theory with $N_f$ flavors of fermions in representation $R$,
\beea
b_1&=&  \frac{11}{3}\, C_2(G) - \frac{4}{3}\,N_f T(R) \label{2loopbeta1}\\
b_2&=& \frac{34}{3}\, [C_2(G)]^2
  -N_f T(R) \left[\frac{20}{3}\, C_2(G) %\right.\nonumber\\&&
  + 4 C_2(R) \right].
\label{2loopbeta2}
\eea
Here $C_2(R)$ is the value of the quadratic Casimir operator in representation $R$ [where $G$ denotes the adjoint representation, so $C_2(G)=N$], while $T(R)$ is the conventional trace normalization.
We are interested in asymptotically free theories, so we demand $b_1>0$ to force an infrared repulsive fixed point at $g=0$.
If $b_2<0$ while $b_1>0$, the two-loop beta function will have a zero~\cite{Caswell:1974gg,Banks:1981nn} at some $g=g_*$.
Whether the theory is a candidate for strongly
interacting dynamics or for conformal behavior at long distance scales depends on whether the complete, nonperturbative $\beta(g^2)$ has a zero.
If it does have such an infrared attractive fixed point (IRFP),
then we have an unparticle theory with a scale-invariant coupling $g_*$ at large distances. If it does not, so that $g$ runs to strong coupling in the infrared, we have 
a technicolor candidate. The technicolor category also includes the possibility of a coupling region where $\beta(g^2)$ approaches zero without crossing it. The coupling would then evolve slowly between widely differing energy scales.
 ``Walking technicolor'' is built on this scenario~\cite{Holdom:1981rm,Yamawaki:1985zg,Appelquist:1997fp}.

In a massive theory, the running coupling $g(q^2)$ is supplemented by the running fermion mass $m(\mu)$.
The counterpart of the beta function is the anomalous dimension $\gamma_m$ of the mass operator $\bar\psi\psi$.
It determines the running of the mass parameter according to
\bee
\mu \frac{\partial m(\mu)}{\partial \mu} = -\gamma_m(g^2) m(\mu).
\ee
If the system is conformal at zero fermion mass $m_q$, then near $m_q=0$ the correlation length $\xi$ scales as
\bee
\xi \sim m_q^{-{1}/{y_m}}
\ee
where $y_m = 1+\gamma_m(g_*)$ is the leading relevant exponent of the system (in the language of critical phenomena).
In lowest order in perturbation theory,
\bee
\gamma_m= \frac{6 C_2(R)}{16\pi^2} g^2.
\ee

In the massless theories used for technicolor, $\gamma_m$ governs the running of the condensate $\svev{\bar\psi\psi}$.
It is thus an important diagnostic for realistic ``extended'' technicolor models. 
Phenomenological constraints on such models require it to have a large, nonperturbative value.

Briefly~\cite{Chivukula:2010tn}, at issue is the dual role of the conventional Higgs condensate in giving masses both to the weak bosons $W,Z$ and to the quarks and leptons.
In technicolor theories, the $W$ mass comes from the techniquark condensate $\svev{\bar\psi\psi}$; this demands that the technicolor scale $\Lambda_{TC}$ be not much above the weak scale, $\Lambda_{TC}\sim 1$~TeV.
The light fermions, on the other hand, get their mass from joining the techniquarks in a multiplet of an extended technicolor (ETC) gauge group, which breaks to the technicolor group at scale $\Lambda_{ETC}$.
The coupling of the techniquark condensate to the light fermions via emission of ETC gauge bosons gives the light fermions their masses,
\bee
m_f\sim\frac{\svev{\bar\psi\psi}}{\Lambda_{ETC}^2}
\sim\frac{\Lambda_{TC}^3}{\Lambda_{ETC}^2}.
\label{mf}
\ee
On the other hand, the exchange of ETC gauge bosons generates flavor-changing neutral currents (FCNC) among the light fermions, with effective vertices $\sim 1/\Lambda_{ETC}^2$.
Suppression of FCNC demands that $\Lambda_{ETC}$ be pushed up beyond 1000~TeV, but then the fermion masses generated by \Eq{mf} come out too light, as small as $m_f\alt1$~MeV.

The solution of this problem in ETC theories lies in the recognition that the two uses of the $\svev{\bar\psi\psi}$ condensate involve its evaluation at two very different energy scales.
The weak boson masses are connected to the condensate at $\Lambda_{TC}$, while the fermion masses are determined by its value at $\Lambda_{ETC}$.
The condensate runs between these two scales according to its anomalous dimension, so that the numerator in \Eq{mf} is really
\bee
  \left. \langle \bar\psi\psi \rangle \right|_{ETC} =
  \left. \langle \bar\psi\psi \rangle \right|_{TC}
  \exp\left[\int_{\Lambda_{TC}}^{\Lambda_{ETC}} \frac{d\mu}{\mu}\,\gamma_m\left(g^2(\mu)\right)\right].
\ee
If $\gamma_m$ is approximately constant, then 
\bee
\left. \langle \bar\psi\psi \rangle \right|_{ETC} \approx
\left. \langle \bar\psi\psi \rangle \right|_{TC}
\left(\frac{\Lambda_{ETC}}{\Lambda_{TC}}\right)^{\gamma_m}.
\ee
Since the ratio $\Lambda_{ETC}/\Lambda_{TC}\sim10^{3}$,
a ``condensate enhancement'' of this order can account for most quark masses---but only if $\gamma_m$ is close to 1.

The walking-technicolor scenario combines a near-zero of the beta function with a large anomalous dimension for the mass.
One envisions a (perhaps) rapid evolution of the coupling away from $g=0$ at some extremely high energy scale, which slows to near-fixed-point behavior at some moderately large coupling $g$ at scale $\Lambda_{ETC}$.
The coupling then runs very slowly until $\Lambda_{TC}$ is reached, even as the techniquark mass runs with a large anomalous dimension $\gamma_m$.
At $\Lambda_{TC}$, the coupling is a bit stronger and creates the techniquark condensate; the techniquarks thereupon decouple and the coupling runs on to strong values, leaving technicolor confined at low energies.

Most of the ingredients in this story---near-fixed-point behavior, condensation, and subsequent confinement---involve nonperturbative physics.
Lattice methods, which have been fairly successful in dealing with the properties of QCD, are a natural approach to investigate them \cite{Nelson:2006zz}.
Every candidate model for new physics begs the two questions:
\begin{enumerate}
\item Is the long distance dynamics of the candidate theory conformal, or confining, or something else?
\item What is the value of $\gamma_m$ in the interesting energy range?
\end{enumerate}

\begin{figure}
\begin{center}
\includegraphics[width=\columnwidth,clip]{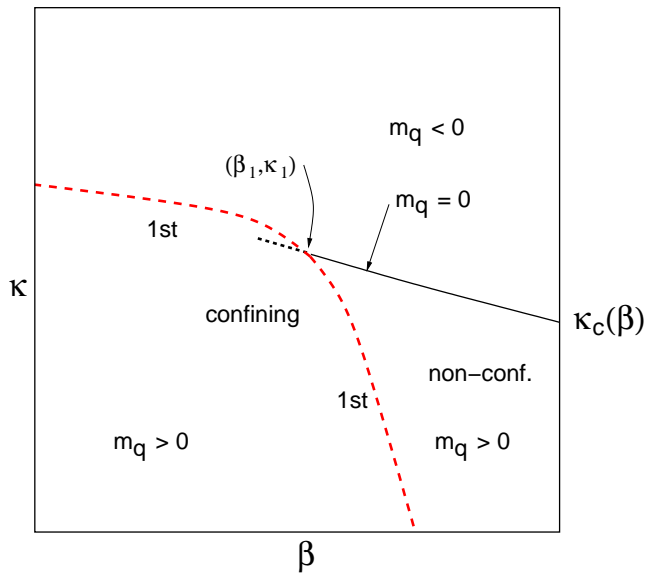}
\end{center}
\caption{Sketched phase diagram of the lattice theory in finite volume, as found for both thin links~\cite{DeGrand:2008kx} and fat links (this paper).
For quantitative information see Figs.~\ref{fig:bktSF} and \ref{fig:bktAP}.
The $\kappa_c(\beta)$ curve (solid), where $m_q=0$, exists only in the non-confining phase; it meets the phase boundary (dashed) at $(\beta_1,\kappa_1)$.
The dotted curve indicates the extension of $\kappa_c(\beta)$ into the confining phase via the metastable non-confining state.
\label{sketch}}
\end{figure}

For some time our program has been to study the gauge theory proposed for ``next-to-minimal'' walking technicolor~\cite{Sannino:2004qp,Hong:2004td, Dietrich:2005jn,Belyaev:2008yj}:
SU(3) gauge fields coupled to two flavors of fermions in the two-index symmetric representation, which is the sextet.
In Ref.~\cite{Shamir:2008pb} we measured the running coupling constant defined in the Schr\"odinger-functional background field method.
We found, with the small lattices at our disposal, that the integrated beta function has a zero, indicating that the theory is in a conformal phase.
In Ref.~\cite{DeGrand:2008kx} we performed spectroscopic studies in the theory with non-zero quark mass. We observed two phases for the system in finite volume: a strong-coupling, confined phase and a weakly coupled, chirally restored phase.
(See Fig.~\ref{sketch}.
The massless $\kappa_c(\beta)$ curve is where our Schr\"odinger functional calculations are done.)
While one might interpret the two phases as low- and high-temperature phases, the phase transition in the massless theory [i.e., at $(\beta_1,\kappa_1)$] appeared not to move as the lattice volume increased; this is in marked contrast to the usual high-temperature phase transition in confining gauge theories.
%%%%%%%%%%%%%%%%%%%%%%%%%%%%%%%%%%%%%%%%%%%%%%%%%%%%%%%%%%%%%%%%%%%%%
%%%%%%%%%%%%%%%%%%%%%%%%%%%%%%%%%%%%%%%%%%%%%%%%%%%%%%%%%%%%%%%%%%%%%

Following these calculations, we changed the lattice action from the usual Wilson--clover action (``thin links'') to an improved action, incorporating normalized hypercubic smearing (``nHYP fat links'') that has shown dramatic reduction of lattice artifacts when used for QCD simulations.
One of us \cite{DeGrand:2009hu} performed a finite-size scaling analysis of correlation lengths in the weak coupling phase, approaching the massless $\kappa_c$ curve.
A small anomalous dimension, $\gamma_m\simeq 0.5$, was observed at two values of the bare parameters.%
\footnote{For other recent work on SU(3) gauge theory with sextet fermions, see~\cite{Fodor:2009ar,Machtey:2009wu,Kogut:2010cz}.}

In this paper we report on a new calculation of the running coupling constant and of the mass anomalous dimension,
using the improved action on lattice volumes larger than in Refs.~\cite{Shamir:2008pb, DeGrand:2008kx}.
The first step is to determine the phase structure of the lattice theory.
As with the original, thin-link action, the improved, fat-link action has a confined phase in which it is not possible to tune the quark mass to zero.
We go beyond Ref.~\cite{DeGrand:2008kx} in studying closely the region where the $\kappa_c$ line intersects the phase boundary, and we find that there is no critical point.
The phase boundary moves up the $\kappa_c$ curve very slowly as the simulation volume grows; it is also affected by the choice of boundary conditions.
As noted in the context of other, similar gauge theories (e.g., \cite{Iwasaki:2003de}), this phase diagram is different from that of QCD and it is hard to see how it can tend to a confining theory in the continuum limit.
Such considerations, however, are far from conclusive.

Turning to the SF calculation, we find that the coupling constant runs more slowly than two-loop perturbation theory would predict.
This is such slow running that we are unable to chain together results from different sets of bare parameters to construct a picture of $g(L)$ running over decades of the length scale $L$.
The range of couplings that we can investigate is limited by the strong-coupling phase transition.
The beta function might  cross zero at the strongest coupling that we can reach, but we cannot make a strong claim to this effect.

Our previous calculation~\cite{Shamir:2008pb}, with the thin-link action, was unable to pursue the $\beta$ function to SF couplings stronger than $g^2\simeq2.5$ because of the transition to the strong-coupling phase.
The infrared fixed point found in that work was at a renormalized coupling $g^2\simeq2.0$ (see Fig.~\ref{fig:DBF} below), which was uncomfortably close to the transition.
With the fat-link action we are able to push the transition to $g^2\simeq 5$, so that the region around $g^2\simeq2.0$ is well insulated from lattice artifacts.
Thus the present calculation is more reliable when it rules out the fixed point found previously.

As part of the SF calculation, we compute the anomalous dimension $\gamma_m$.
Even when one takes into account possible finite-lattice effects,
the result has much smaller uncertainty than the calculation of Ref.~\cite{DeGrand:2009hu} and confirms its conclusion: $\gamma_m$ is small over our observed range of couplings.

The outline of the paper is as follows: In Sec.~\ref{sec:method} we review our lattice action and the techniques we use to measure the beta function and $\gamma_m$.
In Sec.~\ref{sec:SC} we describe our studies of the boundary between the strong- and weak-coupling phases.
Sections~\ref{sec:SF} and~\ref{sec:zp} contain our results for the running coupling constant and mass anomalous dimension.
We summarize the calculation in Sec.~\ref{sec:last}, and place it in context of other lattice calculations.
The appendix contains data that support our determination of the phase boundary as presented in Sec.~\ref{sec:SC}.

%%%%%%%%%%%%%%%%%%%%%%%%%%%%%%%%%%%%%%%%%%%%%%%%%%%%%%%%%%%%%%%%%%%%%
\section{Methodology \label{sec:method}}
%%%%%%%%%%%%%%%%%%%%%%%%%%%%%%%%%%%%%%%%%%%%%%%%%%%%%%%%%%%%%%%%%%%%%

\subsection{Lattice action and simulation}

We study the SU(3) gauge theory with two flavors of dynamical fermions in the sextet representation of the color gauge group.
The lattice action is given by the single-plaquette gauge action and a
Wilson fermion action with added clover term~\cite{Sheikholeslami:1985ij}.
The gauge connections in the fermion action employ the differentiable hypercubic smeared link of Ref.~\cite{Hasenfratz:2007rf}, from which the symmetric-representation gauge connection for the fermion operator is constructed. 
The parameters that are inputs to the simulation are the bare gauge coupling $\beta=6/g_0^2$  and the fermion hopping parameter $\kappa$, related to the bare mass $m_0$ by $\kappa=(8+2m_0)^{-1}$.
Unlike our earlier calculation with the thin-link fermion action, no tadpole improvement is necessary here and thus we set the clover coefficient to its tree-level value (i.e., unity).
The smearing parameters for the links are the same as in
Ref.~\cite{Hasenfratz:2007rf}: $\alpha_1=0.75$, $\alpha_2=0.6$, $\alpha_3=0.3$.

The molecular dynamics integration is accelerated with an additional heavy pseudo-fermion field as suggested by Hasenbusch~\cite{Hasenbusch:2001ne}, multiple time scales~\cite{Urbach:2005ji},
and a second-order Omelyan integrator~\cite{Takaishi:2005tz}.
Lattice sizes range from $6^4$ to $16^4$ sites; some data for the phase diagram were obtained with lattices of $12\times6^3$ sites.

\subsection{Why fat links?}

The simulations we reported in Ref.~\cite{Shamir:2008pb} were performed using the usual clover action~\cite{Sheikholeslami:1985ij} in which the coefficient $c_{SW}$ of the clover term was
adjusted via tadpole improvement.
When we began the present set of simulations, we were faced with a choice: either to continue using the same action  and simply to push to larger lattice volumes, or to attempt as well to improve the action further.
We chose the latter course.

Improvement criteria are basically identical for any asymptotically free theory.
As one tunes the bare coupling toward zero, the lattice theory should have an expansion in powers of the lattice spacing $a$.
The dimension-four terms in this expansion are identical to those of a continuum action, while the higher-dimensional terms, which are multiplied by positive powers of the lattice spacing, correspond to irrelevant operators.
The inclusion of a clover term, for instance, allows the fermion action to be free of $O(a)$ artifacts if the clover term is appropriately chosen. 
In general, the particular choice of lattice discretization is formally irrelevant.
In practice, however, it will make a big difference in terms of eliminating lattice artifacts.
We chose hypercubic smeared links because of our positive experience with them in QCD simulations.

Fermionic actions with hypercubic smeared links have a number of favorable features, which we enumerate in the context of the present simulation:
\begin{enumerate}
\item The gauge fields seen by the fermions are smoother with a smeared link than they are with a thin link.
The plaquette (in the fermions' color representation) is a rough indicator of this.
For example, at $(\beta=4.4,\kappa=0.1351)$ the thin-link (fundamental) plaquette has a 
value of about 0.43 (normalized to a maximum of 1) while the sextet-representation smeared-link plaquette is about 0.78 (normalized likewise).
Thus, even at this strong (bare) coupling, one can imagine
expanding the link variable $U_\mu \sim 1 + ig A_\mu +\cdots$ in order to recover the continuum action.
%(The gauge part of the action uses thin links, so it is more susceptible to lattice artifacts.)
\item The value of $\kappa_c(\beta)$, at which the quark mass (see below) vanishes, is closer to the free-field value of 1/8 with smeared links than it is in our original thin link calculation.
Even at the boundary of the strong-coupling phase we find
$\kappa_c \alt 0.136$, versus a range of 0.15--0.17 for the entire range of couplings studied in the thin-link theory in Ref.~\cite{Shamir:2008pb}.
\item Tests with quenched fundamental fermions \cite{Hoffmann:2007nm}
reveal that the nonperturbatively improved clover coefficient is very close to its tree level value at fairly large lattice spacings.
Preliminary results show this to be the case in the present theory as well~\cite{evgeny}.
This is our justification for setting $c_{SW}=1$ here.
\item Finite renormalization factors for (partially) conserved currents are much closer to unity than for the thin-link clover action. 
This is easily checked in one-loop perturbation theory, where the vector and axial-vector lattice-to-continuum renormalization factors are $Z= 1 + g^2 C_2(R) c$, where $c$ is a pure number, a lattice integral~\cite{DeGrand:2002vu}.
\end{enumerate}

While the formal arguments for improvement via smearing may be called into question outside the weak-coupling limit, the practical observations listed above justify its application and show why results from the smeared theory are more reliable than those obtained without smearing.
Nonetheless, asymptotic freedom does underlie the whole philosophy.
If an infrared fixed point were to be found, the theory on the strong-coupling side of the fixed point would not be asymptotically free.
Little is known {\em a priori\/} about such a strong-coupling theory.

As mentioned in the introduction, we find in the present study that the use of hypercubic smeared links allows us to reach a much larger value of the Schr\"odinger functional coupling $g^2$ than we could get to in Ref.~\cite{Shamir:2008pb}, $g^2\simeq 5$
versus about 2.5, before encountering the strong-coupling phase transition.
Thus the region $g^2\simeq2.0$, where the thin-link calculation indicated an infrared fixed point, can be studied more reliably.

\subsection{Schr\"odinger functional method and the running coupling}

The Schr\"odinger functional (SF)~\cite{Luscher:1992an,Luscher:1993gh,Sint:1995ch,Jansen:1998mx} method is an implementation of the background field method that is especially suited for a lattice calculation. 
Taking the simulation volume to be a 4-cube of dimension $L$, one imposes fixed boundary conditions on the gauge field at $t=0$ and $t=L$ while imposing periodic boundary conditions in the spatial directions.
The classical field that minimizes the Yang--Mills action
subject to the fixed boundary conditions is a background color-electric field.
By construction the only distance scale that characterizes the background field is $L$, so
the $n$-loop effective action $\Gamma\equiv-\log Z$ gives the running coupling via
\bee
\label{Gamma}
\Gamma = g(L)^{-2} S_{YM}^{cl} , 
\ee
where
\bee
\label{SYM}
S_{YM}^{cl} = \int d^4x\, F^2_{\mu\nu}
\ee
is the classical action of the background field.
When $\Gamma$ is calculated non-perturbatively, \Eq{Gamma} gives a non-perturbative definition of the running coupling at scale $L$.

Since in a lattice calculation one cannot calculate $\Gamma$ itself,
one differentiates \Eq{Gamma} with respect to some parameter $\eta$ in the boundary conditions.
Thus
\bee
\left.\frac{\partial \G}{\partial\eta} \right|_{\eta=0}
= \frac{K}{g^2(L)}\,,
\qquad\text{where}\ 
K \equiv \left.\frac{\partial S_{YM}^{cl}}{\partial\eta} \right|_{\eta=0} .
\label{defg}
\ee
The derivative of $\Gamma$ gives an observable quantity, while $K$ is just a number \cite{Luscher:1993gh}.
We choose boundary fields as described in Ref.~\cite{Jansen:1998mx};
for these boundary values the coefficient $K\simeq37.7$.
The observable,
\bee
  \left.\frac{\partial \Gamma}{\partial\eta} \right|_{\eta=0}
  =
  \left.\svev{\frac{\partial S_{YM}}{\partial\eta}
  -\tr \left( \frac{1}{D_F^\dagger}\;
        \frac{\partial (D_F^\dagger D_F)}{\partial\eta}\;
            \frac{1}{D_F} \right)}\right|_{\eta=0},
            \label{deta}
\ee
is a particular expectation value of the gauge fields and the Dirac operator $D_F$.
The parameter $\eta$ enters linearly into phase angles so the derivatives on the right-hand side of \Eq{deta} are implemented by putting the appropriate fixed values in place of the boundary links \cite{Luscher:1993gh}.
We also impose twisted spatial boundary conditions on the fermion fields
as suggested in Ref.~\cite{Sint:1995ch},
$\psi(x+L)=\exp(i\theta)\psi(x)$, with $\theta=\pi/5$ on all three
axes \cite{DellaMorte:2004bc}.

The observable (\ref{deta}) is quite noisy and requires long simulation runs, as shown in Table~\ref{table:runs}.
The acceptance in the hybrid Monte Carlo simulations was kept at 80--90\%, except at the strongest couplings for $L/a=16$ where we were forced to use short time steps even to reach acceptances as low as 40\%.
%%%%%%%%%%%%%%%%%%%%%%%%%%%%%%%%%%%%%%%%%%%%%%%%%%%%%%%%%%%%%%%%%%%%%
\begin{table}
\caption{Number of hybrid Monte Carlo trajectories (of unit length) needed to produce the Schr\"odinger functional coupling $g^2$ at the bare couplings $(\beta, \kappa_c)$, for the lattice sizes $L$ used in this study.}
\begin{center}
\begin{ruledtabular}
\begin{tabular}{ccrrrr}
$\beta$ & $\kappa_c$ &\multicolumn{4}{c}{trajectories}\\
\cline{3-6}
&&              $L=6a$    & $L=8a$    & $L=12a$   & $L=16a$  \\
\hline
5.8 & 0.12835 &  8 000 &  2 300 & 10 000 & \hfil--  \\
5.4 & 0.12920 &  8 000 &  3 050 &  9 600 & \hfil--  \\
5.0 & 0.13062 & 13 000 &  6 030 &  8 800 & \hfil--  \\
4.8 & 0.13173 & 45 000 &  5 250 & %9200(B)+7500(T)
                                  16 700 & \hfil--  \\
4.6 & 0.13320 & 74 000 & 13 400 & %6518(B)+4758(T)
                                  11 300 & 22 400 \\
4.4 & 0.13510 & 67 200 & 22 900 & 11 300 & 29 200 \\
4.3 & 0.13617 & 14 800 & 15 600  & 8 000 &  6 750 \\
\end{tabular}
\end{ruledtabular}
\end{center}
\label{table:runs}
\end{table}
%%%%%%%%%%%%%%%%%%%%%%%%%%%%%%%%%%%%%%%%%%%%%%%%%%%%%%%%%%%%%%%%%%%%%

\subsection{Anomalous dimension}
The mass anomalous dimension is determined from the volume dependence of the renormalization factor $Z_P$ of the isovector pseudoscalar density $P^a=\bar\psi\gamma_5(\tau^a/2)\psi$.
(The latter is related by a chiral rotation to $\bar\psi\psi$, which is the object of interest.)
It is computed from two correlators via~\cite{Sint:1998iq,Capitani:1998mq,DellaMorte:2005kg,Bursa:2009we}
\bee
Z_P = \frac {c \sqrt{f_1}}{f_P(L/2)}.
\label{eq:ZP}
\ee
$f_P$ is the propagator from the $t=0$ boundary to a point pseudoscalar operator at time $x_0$,
\beea
  f_P(x_0)&=&-\frac{1}{3}\sum_a \int d^3y\, d^3z\,  \left\langle
  \overline{\psi}(x_0)\gamma_5\frac{\tau^a}{2}\psi(x_0)\right.\nonumber\\
  &&\times\left.\overline{\zeta}(y)\gamma_5\frac{\tau^a}{2}\zeta(z)
  \right\rangle,
  \label{eq:fPdef}
\eea
and we take $x_0=L/2$; here $\zeta$ and $\bar\zeta$ are gauge-invariant wall sources at $t=a$, meaning one lattice layer away from the $t=0$ boundary.
The normalization of the wall source is removed by the $f_1$ factor, which is the boundary-to-boundary correlator,
\beea
  f_1&=&-\frac{1}{3L^6}\sum_a \int d^3u\, d^3v\, d^3y\, d^3z\,
  \left\langle
  \overline{\zeta}^\prime(u)\gamma_5 \frac{\tau^a}{2}{\zeta}^\prime(v)
  \right.\nonumber\\
  &&\times\left.\overline{\zeta}(y)\gamma_5\frac{\tau^a}{2}\zeta(z)
  \right\rangle,
  \label{eq:f1def}
\eea
where $\zeta'$ and $\bar\zeta'$ are wall sources at $t=L-a$.
The constant $c$ allows imposing a volume-independent normalization condition in the weak-coupling limit.

In SF calculations for QCD, correlators such as (\ref{eq:fPdef}) and~(\ref{eq:f1def}) are usually computed with the
spatial link matrices at $t=0$ and $t=L$ set to unity.
This is because it is desired to compute $Z_P$ with an absolute normalization (to fix the physical value of a quark mass, for example). 
Correspondingly, the constant $c$ that normalizes $Z_P$ has only been calculated for these boundary conditions.
Since our interest, however, is in how $Z_P$ runs with the scale $L$, we will calculate only ratios of values of $Z_P$ for different $L$ at fixed lattice couplings.
Because of this, the overall normalization of $Z_P$ is irrelevant.
We are thus free to ignore $c$ and also to use the same boundary conditions for the calculation of $Z_P$ as for the simulations which generated the data for the SF coupling.
Thus the data for $Z_P$ came from the same configurations as the coupling calculation, effectively giving $Z_P$ for free.
We set $c=1/\sqrt2$ in tabulating $Z_P$ below.

\subsection{Quark mass}
We studied the massless theory by fixing $\kappa=\kappa_c(\beta)$,
the point at which the quark mass $m_q$ vanishes for each $\beta$.
We define $m_q$ using the unimproved axial Ward identity (AWI),
\bee
\partial_t \sum_\bx \svev{A_0^a(\bx,t)\co^a} = 2m_q \sum_\bx \svev{ P^a(\bx,t)\co^a}.
\label{eq:AWI}
\ee
where $A_0^a=\bar \psi \gamma_0\gamma_5 (\tau^a/2)\psi$ and $\co^a$ could
be any source. We follow the usual SF procedure and take the source to be the gauge-invariant wall source at $t=a$ as in \Eq{eq:fPdef}.
The correlation functions in \Eq{eq:AWI} are then measured at $t=L/2$, the midpoint of the lattice.
The derivative is taken as the symmetric difference, $\partial_\mu f(x)=[f(x+\hat\mu a) - f(x-\hat\mu a)]/(2a)$.

%%%%%%%%%%%%%%%%%%%%%%%%%%%%%%%%%%%%%%%%%%%%%%%%%%%%%%%%%%%%%%%%%%%%%
\section{Strong coupling \label{sec:SC}}
%%%%%%%%%%%%%%%%%%%%%%%%%%%%%%%%%%%%%%%%%%%%%%%%%%%%%%%%%%%%%%%%%%%%%
\subsection{First-order phase boundary}

As shown in Fig.~\ref{sketch}, both the thin-link and the fat-link lattice theories show a first-order phase transition separating a low-$\beta$ (strong bare coupling) phase from a high-$\beta$ (weak bare coupling) phase.
Our work on the thin-link~\cite{DeGrand:2008kx} and fat-link~\cite{DeGrand:2009hu} theories showed that:
\begin{itemize}
\item The low-$\beta$ phase is confining, as revealed by the heavy-quark potential (for sources in the fundamental representation).
\item The high-$\beta$ phase is chirally restored, as revealed by parity doubling between the scalar and pseudoscalar masses, and also between the vector and axial-vector masses.
The string tension is unobservably small in this phase.
\item In the high-$\beta$ phase, all screening masses and the pseudoscalar decay constant $f_\pi$ fall towards zero as $m_q\rightarrow 0^+$, until the correlation length $\xi$ ($\equiv 1/m_\pi$) approaches $L$, the length of the lattice in a direction transverse to the direction in which the correlator is measured.
Then $\xi$ plateaus at a value proportional to $L$. This behavior superficially resembles the usual finite-size scaling for a critical system in finite volume, where $1/L$ plays the role of a relevant perturbation~\cite{Cardy:1996xt}.
(This behavior was exploited in Ref.~\cite{DeGrand:2009hu}.)
\end{itemize}

While the two phases appear for both SF boundary conditions and for the usual thermal [i.e., (anti-)periodic temporal] boundary conditions, there are important differences between the two cases.
For thermal boundary conditions (BC) the Polyakov loop (in the fundamental representation) can be used as usual to distinguish between the two phases.
In the strong-coupling phase it is near zero, real and slightly negative.
(It will be recalled that the fermion action with thermal BC breaks the global Z(3) symmetry of the gauge action.)
In the weak-coupling phase the Polyakov loop $P$ orders along one of the Z(3) center elements.
The state with $\svev P$ real and positive is the stable one, while the states which order along the other directions, $\svev P \sim \exp(\pm2\pi i/3)$, are metastable.\footnote{
We observed tunneling among these states on lattices with volume $8^4$. 
See Ref.~\cite{Machtey:2009wu} for a study of the thin-link theory, and compare~\cite{Kogut:2010cz}.}
This allows the characterization of the phase transition as a finite-temperature confinement transition, as far as is possible in a theory with dynamical fermions.

With SF boundary conditions (SFBC), on the other hand, the fermion action does {\em not\/} break the Z(3) center symmetry---the Z(3) symmetry is exact, just as in the pure gauge theory.
As it turns out, the Z(3) symmetry is spontaneously broken in the
weak-coupling phase; it is unbroken in the strong-coupling phase, so
that <P>=0 there, much the same as in a pure gauge theory with thermal
BC.
Of course, a gauge theory with SFBC has no interpretation as a true finite-temperature system.

In both cases, the entire phase boundary is strongly first-order, as seen in discontinuities in the plaquette, in the AWI quark mass, in the ordering of the Polyakov loop, and, most of all, in hysteresis.
Simulations initialized in one phase but run at parameters in the other phase can run for hundreds of molecular dynamics time steps without tunneling.
We present plots of the mean plaquette and the AWI quark mass in the appendix.

\begin{figure}
\begin{center}
\includegraphics[width=\columnwidth,clip]{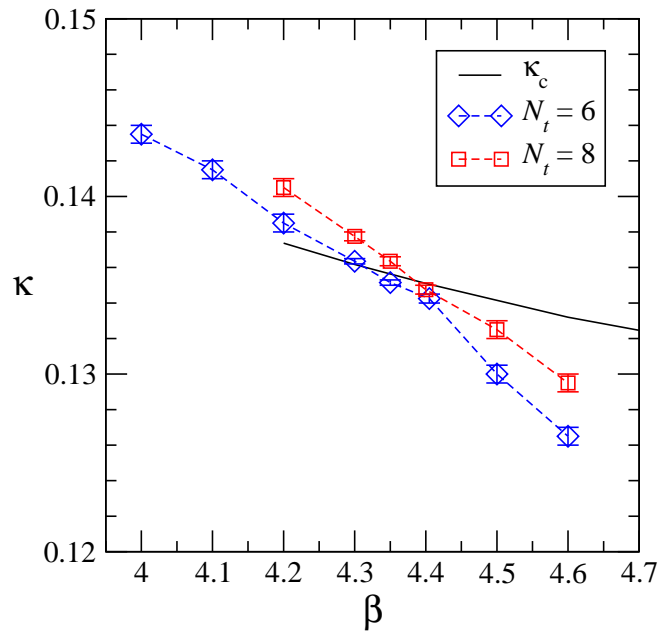}
\end{center}
\caption{Phase diagram with SF boundary conditions. The curve is 
$\kappa_c(\beta)$.
Symbols show the location of the phase boundary between the strong coupling phase at lower $\beta$ and the weak coupling phase at higher $\beta$.
Diamonds are from $12\times6^3$ lattices ($\beta=4.0$--4.2) and $6^4$ lattices ($\beta=4.3$--4.6) while squares are from $8^4$ lattices. 
\label{fig:bktSF}}
\end{figure}

\begin{figure}
\begin{center}
\includegraphics[width=\columnwidth,clip]{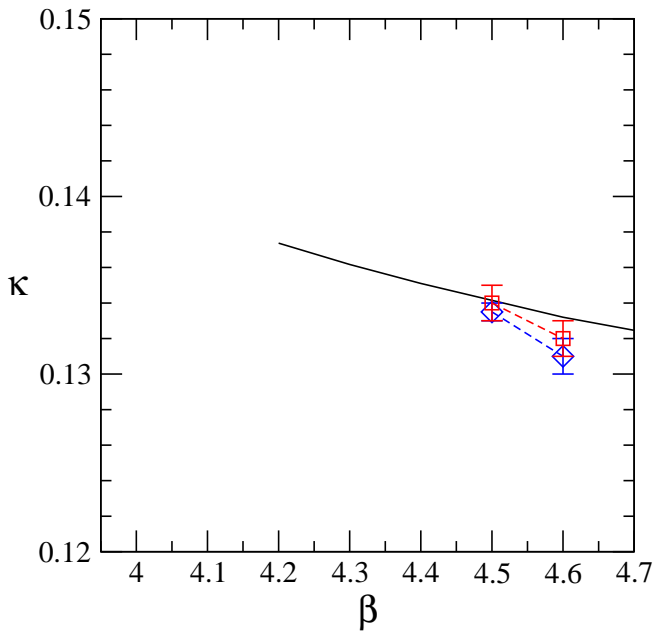}
\end{center}
\caption{Phase diagram with thermal boundary conditions. 
The curve is $\kappa_c(\beta)$. 
Symbols show the location of the phase boundary.
Diamonds are from $12^3\times 6$ volumes while squares are from $12^3\times 8$ volumes.
\label{fig:bktAP}}
\end{figure}

We located the phase boundary by performing simulations with mixed starts---one half of the initial gauge configuration is taken from an equilibrium configuration at some $(\beta, \kappa)$ in one phase, while the other half is from an equilibrium lattice at $(\beta',\kappa')$ in the other phase.
We then watch the system equilibrate.
This can be quite expensive, as it requires tiny time steps to avoid rejection of the molecular dynamics trajectory in the HMC algorithm.
We have thus only done these tests for lattices of temporal extent $N_t=6$ and~8: $6^4$, $12\times6^3$, and $8^4$ lattices with SFBC, and $12^3\times 6$ and $12^3\times 8$ lattices with thermal BC.
The location of the phase boundary depends on the choice of BC as well as on $N_t$.
We show the region of the intersection of the phase boundary with the $\kappa_c$ curve for the various cases in Figs.~\ref{fig:bktSF} and~\ref{fig:bktAP}.

%%%%%%%%%%%%%%%%%%%%%%%%%%%%%%%%%%%%%%%%%%%%%%%%%%%%%%%%%%%%%%%%%%%%%
%%%%%%%%%%%%%%%%%%%%%%%%%%%%%%%%%%%%%%%%%%%%%%%%%%%%%%%%%%%%%%%%%%%%%
%%%%%%%%%%%%%%%%%%%%%%%%%%%%%%%%%%%%%%%%%%%%%%%%%%%%%%%%%%%%%%%%%%%%%

It should not be surprising that the phase boundary shifts when SFBC are replaced by thermal.
For given $N_t$, thermal lattices have $N_t$ spacelike layers of dynamical links, while SF lattices have only $N_t-1$.
The smaller number of dynamical degrees of freedom means that the SF lattice is effectively at a higher ``temperature'' than the thermal lattice at the same $(\beta,\kappa)$.
Indeed, comparison of Figs.~\ref{fig:bktSF} and~\ref{fig:bktAP} shows that the phase boundary for the SF theory lies below and to the left of that of the thermal theory with the same $N_t$.
This behavior is familiar in thermal gauge theories when $N_t$ is varied.
We repeat, however, that the SF theory is {\em not\/} a finite-temperature theory and that the phase transition in the SF theory is {\em not\/} a finite-temperature transition.

We performed SF calculations on the $\kappa_c$ line, defined by $m_q=0$.
In the weak coupling phase it is quite easy to find $\kappa_c$, and its volume dependence is small.
When $\beta$ is less than $\beta_1$, the gauge coupling where the $\kappa_c$ line meets the first order line,
the quark mass is never zero; it changes sign discontinuously when $\kappa$ crosses the phase boundary (see the appendix).
This means that there is no massless theory to the left of $\beta_1$.
The $\kappa_c$ curve can, however, be extended leftward into the strong-coupling phase by simulating in the {\em metastable\/} state, which is continuously connected to the weak-coupling phase.
It is in fact possible to do long runs in this state without tunneling out if it.
We can calculate $m_q$, which does cross zero and thus gives us $\kappa_c$; we can also calculate $g^2(L)$ and $Z_P(L)$.
As we will show, these observables remain very similar to their values in the region of bare parameter space where the weak-coupling phase is the true vacuum.

There is another metastable extension of the $\kappa_c$ line, found by following a confining state into the non-confining phase.
We have not explored this because simulation in confining states is much more difficult.

\subsection{Shift of the phase boundary with lattice size}

For $\kappa<\kappa_c$ the phase boundary shifts to the right as $N_t$ is increased.
For sufficiently small $\kappa$ this is to be expected since this is the scaling behavior typical of the finite-temperature transition when the fermions are massive.
It turns out to be true for $\kappa>\kappa_c$ as well.
Close to $(\beta_1,\kappa_1)$, this shift is very small.
One interpretation of the lack of motion of the phase boundary here is that, close to this point, the transition is a bulk transition, weakly affected by the finite size of the lattice.
The alternative explanation is that (at least for thermal boundary conditions) the transition is still a finite
temperature transition, but that it is moving very slowly. 

To appreciate the distinction, we have to make a comparison to theory.
 From asymptotic freedom we expect the bare coupling $g_0$ at a physical finite-temperature transition to scale with the two-loop formula,
\bee
  \frac{1}{g_0^2(a_2)} = \frac{1}{g_0^2(a_1)} + \hat b_1 x
  + \frac{\hat b_2}{\hat b_1} \log\left(1+\hat b_1 g_0^2(a_1) x\right).
\label{eq:twoloopshift}
\ee
where [see Eqs.~(\ref{2loopbeta})--(\ref{2loopbeta2})] 
$x=\log(a_1/a_2)^2$, 
$\hat b_1=b_1/(16\pi^2)\simeq0.027$, and 
$\hat b_2 = b_2/(16\pi^2)^2\simeq-0.003$.
For $N_t=6$ and~8 the two lattice spacings $a_{1,2}$ are $(6T^*)^{-1}$ and $(8T^*)^{-1}$, respectively, where the transition temperature $T^*$ gives the scale.
Using \Eq{eq:twoloopshift} for the shift in the bare coupling
$\beta=6/g_0^2$, we would predict $\Delta\beta\simeq 0.08$.
In the appendix we show that the true shift near $\kappa_c$ is somewhat smaller than this.
%So either the phase boundary is really standing still, or it is shifting at a rate slower than predicted by perturbation theory.

It is instructive to compare this to quenched QCD, where the deconfinement transition is physical.
In that case its shift, when comparing $N_t=6$ and $N_t=8$ \cite{Kennedy:1984dk,Gottlieb:1985ug},
is similarly smaller than perturbation theory predicts:
The observed shift is $\Delta\beta\simeq0.14$, while the two-loop prediction is 0.24.
For sextet QCD, the shortfall in the observed shift is the same or less.

The size of the shift in $\beta_1$ carries profound implications for the continuum physics of this gauge theory.
If the first-order transition at $(\beta_1,\kappa_1)$ comes to a halt when $L/a$ is large, then there is no massless continuum limit in which the theory shows confinement and chiral symmetry breakdown.
This is because the only place where $m_q=0$ is on the $\kappa_c$ line, which is entirely in the non-confining phase.
One can take a continuum limit along this line by tuning $\beta\to\infty$, but the infrared physics, for any $L/a$, will always be that of a conformal theory with anomalous dimensions.

In order for the massless continuum theory to display confinement, then, the point $(\beta_1,\kappa_1)$ must move towards large $\beta$ as $L/a$ grows.
One would tune $\beta\to\infty$ while keeping $\beta<\beta_1(L)$.
This is not enough, however.
In lattice QCD one maintains confinement and $m_q=0$ at zero temperature by tuning to $\kappa_c$ before taking $\beta\to\infty$ [while maintaining $\beta<\beta_1(L)$].
This is possible in QCD if the $\kappa_c$ line represents a continuous phase transition between the confining phase at $\kappa<\kappa_c$ and the Aoki phase at $\kappa>\kappa_c$.
In our theory, however, the confining phase ends at the first-order boundary and there is no place where $m_q=0$ for $\beta<\beta_1$.

Thus the best one can do at a given $L$ in this scenario is to minimize $m_q$ by tuning the couplings to $[\beta_1(L),\kappa_1(L)]$ from below.
This is {\em not\/} a massless theory.
The massless theory with confinement can only be recovered in the continuum limit; this will happen only if the discontinuity in $m_q$ at $[\beta_1(L),\kappa_1(L)]$ goes to zero as $L/a\to\infty$, which is a possibility but by no means assured.%
\footnote{In fact this is the current state of QCD with Wilson fermions.
The absence of the Aoki phase for large $\beta$ means that there is no theory with $m_q=0$ for finite lattice spacing~\cite{Farchioni:2004us}.
One hopes that the minimal value of $m_q$ at the boundary of the confining phase will go to zero in the continuum limit, i.e., on sufficiently large lattices.}

The observed dependence of $\beta_1$ on $L/a$ is related to our results for the SF beta function (see below).
The derivative $d\beta_1/d(\log L)$ defines a beta function.
The beta function we measure in this way is smaller than its two-loop value near the current value of $\beta_1$.
Our result for the SF beta function behaves similarly.
We find that it is smaller than the two-loop beta function, and in fact we cannot tell whether it actually possesses a zero near $\beta_1$.
In both cases, a zero in the beta function would imply conformal IR physics; a near-zero would imply slow running, and even walking, but confinement in the end.

%%%%%%%%%%%%%%%%%%%%%%%%%%%%%%%%%%%%%%%%%%%%%%%%%%%%%%%%%%%%%%%%%%%%%
\section{Running gauge coupling \label{sec:SF}}
%%%%%%%%%%%%%%%%%%%%%%%%%%%%%%%%%%%%%%%%%%%%%%%%%%%%%%%%%%%%%%%%%%%%%

\begin{figure}
\begin{center}
\includegraphics[width=\columnwidth,clip]{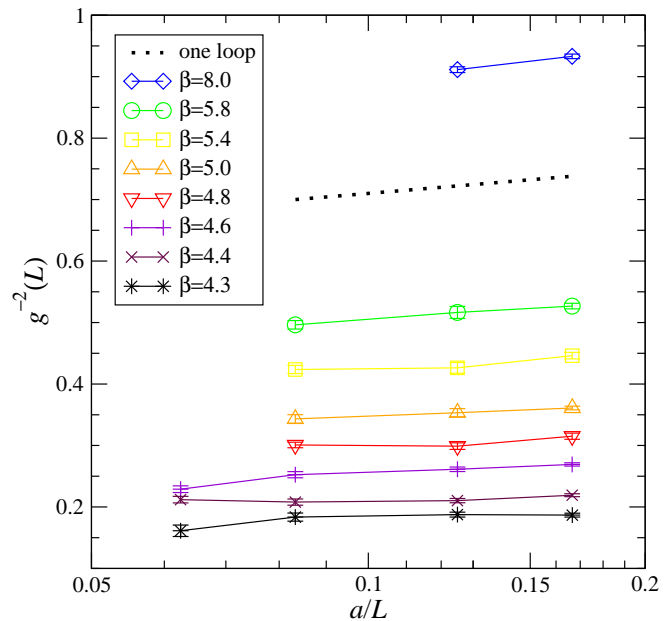}
\end{center}
\caption{SF coupling $1/g^2$ vs.~$a/L$.
The dotted line shows the expected slope from one-loop running.
The data at $\beta=4.3$ were taken in the metastable weak-coupling state.
\label{fig:1g2vsl}}
\end{figure}

Turning now to the SF coupling, we list our results for $g^2(L)$ in Table~\ref{tab:g2table} and plot them in Fig.~\ref{fig:1g2vsl}.
%%%%%%%%%%%%%%%%%%%%%%%%%%%%%%%%%%%%%%%%%%%%%%%%%%%%%%%%%%%%%%%%%%%%%
\begin{table}
\caption{Schr\"odinger functional couplings $g^2$ evaluated at the bare coupling $(\beta, \kappa_c)$ for lattice size $L$.}
\begin{center}
\begin{ruledtabular}
\begin{tabular}{ccllll}
$\beta$ & $\kappa_c$ &\multicolumn{4}{c}{$g^2$}\\
\cline{3-6}
&&              $L=6a$    & $L=8a$    & $L=12a$   & $L=16a$  \\
\hline
5.8 & 0.12835 & 1.898(16) & 1.936(48) & 2.015(28) & \hfil--  \\
5.4 & 0.12920 & 2.241(27) & 2.346(48) & 2.360(37) & \hfil--  \\
5.0 & 0.13062 & 2.770(22) & 2.830(53) & 2.913(59) & \hfil--  \\
4.8 & 0.13173 & 3.173(51) & 3.345(59) & 3.324(50) & \hfil--  \\
4.6 & 0.13320 & 3.715(37) & 3.827(54) & 3.960(79) & 4.37(11) \\
4.4 & 0.13510 & 4.564(50) & 4.755(77) & 4.81(12)  & 4.72(12) \\
4.3 & 0.13617 & 5.355(91) & 5.33(11)  & 5.45(20)  & 6.20(36) \\
\end{tabular}
\end{ruledtabular}
\end{center}
\label{tab:g2table}
\end{table}
%%%%%%%%%%%%%%%%%%%%%%%%%%%%%%%%%%%%%%%%%%%%%%%%%%%%%%%%%%%%%%%%%%%%%
%%%%%%%%%%%%%%%%%%%%%%%%%%%%%%%%%%%%%%%%%%%%%%%%%%%%%%%%%%%%%%%%%%%%%
We compare to the one-loop formula,
\bee
\frac{1}{g^2(L)} = -\frac{2b_1}{16\pi^2}\log L + \text{constant} ,
\label{1loop}
\ee
where $b_1=13/3$, by plotting \Eq{1loop} as the dotted line in the figure.
At $\beta=8$ we find that the coupling runs with the slope of the one-loop result, but $g^2$ runs more slowly at all stronger bare couplings.
For each $\beta$, the change in the coupling over the widest range of $L$ for which we have data is never more than about 15 per cent.
This should be compared to the case of QCD with $N_f=2$, where $b_1=29/3$ is more than twice as large and where in nonperturbative studies the coupling always runs {\em faster\/} than perturbation theory predicts (see Ref.~\cite{DellaMorte:2004bc}).
In an ordinary QCD simulation, one wishes to simulate at bare parameter values where the theory is weakly interacting at short distance, so that one can use perturbation theory to match lattice-regulated
matrix elements to their continuum-regulated counterparts.
One makes the lattice large enough so that the system becomes strongly interacting at long distance, so that the simulation captures the physics of confinement.
Satisfying both conditions does not seem to be possible for sextet QCD.

Note also that there does not seem to be any value of the bare coupling at which $g^2(L)$ clearly decreases as $L/a$ increases.
This means that we cannot argue for the existence of an IRFP.

\begin{figure}
\begin{center}
\includegraphics[width=\columnwidth,clip]{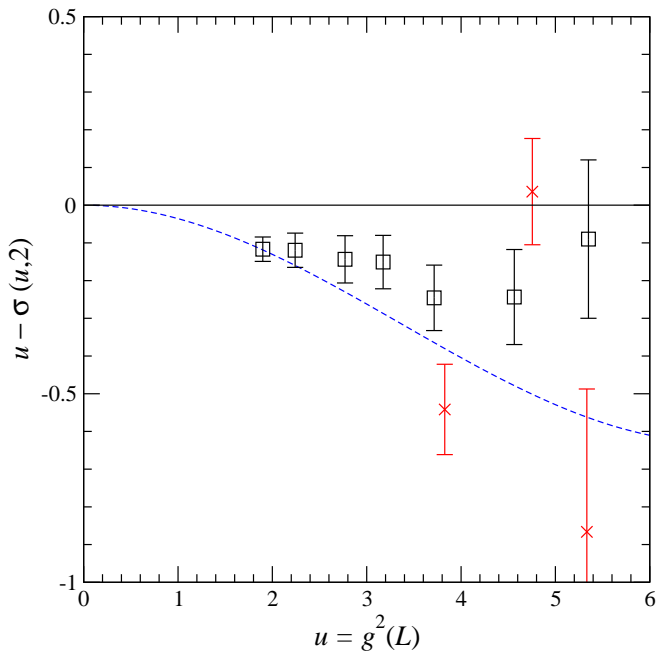}
\end{center}
\caption{Change in coupling under a scale factor $s=2$.
Squares show a comparison of $L=6a\to12a$ while crosses are from $L=8a\to16a$.
The rightmost point in each set was measured in the metastable state at $\beta=4.3$.
The curve is the result of integrating the two-loop perturbative formula.
Here and in other figures where $g^2$ or $1/g^2$ is the abscissa, horizontal error bars, if not drawn, are smaller than the plotting symbols.
\label{fig:betag}}
\end{figure}

\begin{figure}
\begin{center}
\includegraphics[width=\columnwidth,clip]{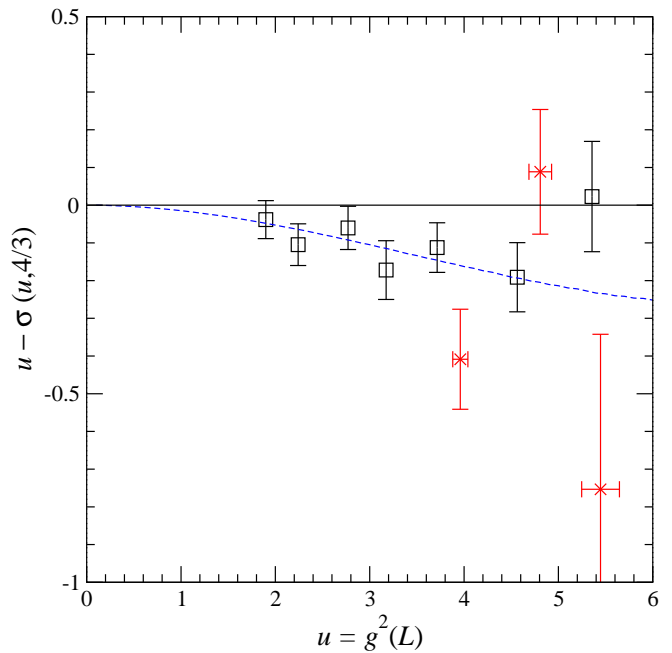}
\end{center}
\caption{Change in coupling under a scale factor $s=4/3$.
Squares show a comparison of $L=6a\to8a$ while crosses are from $L=12a\to16a$.
The rightmost point in each set was measured in the metastable state at $\beta=4.3$.
The curve is the result of integrating the two-loop perturbative formula. 
\label{fig:betag43}}
\end{figure}

From Fig.~\ref{fig:1g2vsl} we derive finite-lattice approximations to the step-scaling function (SSF), conventionally defined by comparing $g^2$ measured on two lattices, viz.,
\bee
\sigma(u,s)=g^2(sL),
\label{SSF}
\ee
where $u=g^2(L)$ and $s$ is the scale factor between the two lattices.
Thus $\sigma(u,s)-u$ is the change in the SF coupling when the lattice IR scale is changed by a factor of $s$.
For scale factor $s=2$ we can compare lattices with $L=6a$ and~$12a$ as well as lattices with $L=8a$ and~$16a$;
for $s=4/3$ we compare $L=6a$ and~$8a$ and also $L=12a$ and~$16a$.
We plot $u-\sigma(u,s)$, which parallels the usual continuum beta function,%
\footnote{For instance, a negative value for $u-\sigma(u,s)$ means asymptotic freedom.}
for these scale factors in Figs.~\ref{fig:betag} and~\ref{fig:betag43}. 
%%%%%%%%%%%%%%%%%%%%%%%%%%%%%%%%%%%%%%%%%%%%%%%%%%%%%%%%%%%%%%%%%%%%%
%%%%%%%%%%%%%%%%%%%%%%%%%%%%%%%%%%%%%%%%%%%%%%%%%%%%%%%%%%%%%%%%%%%%%
%%%%%%%%%%%%%%%%%%%%%%%%%%%%%%%%%%%%%%%%%%%%%%%%%%%%%%%%%%%%%%%%%%%%%
The curves come from integrating the two-loop beta function from $L$ to $sL$ [cf.~\Eq{eq:twoloopshift}].

We see in Fig.~\ref{fig:betag} that the SSF from the $L=6a\to12a$ comparison reflects running that is consistently slower than the integrated two-loop beta function.
It is possible that the rightmost point indicates a fixed point (where we would have $u-\sigma(u)=0$), but the error bar is large; also this point was measured in a metastable state, so its interpretation is unclear.
As for the $L=8a\to16a$ SSF, the most we can say is that it is not inconsistent with that from the smaller lattices.
The SSF for scale factor $s=4/3$ (Fig.~\ref{fig:betag43}) tells a similar story.

\begin{figure}
\begin{center}
\includegraphics[width=\columnwidth,clip]{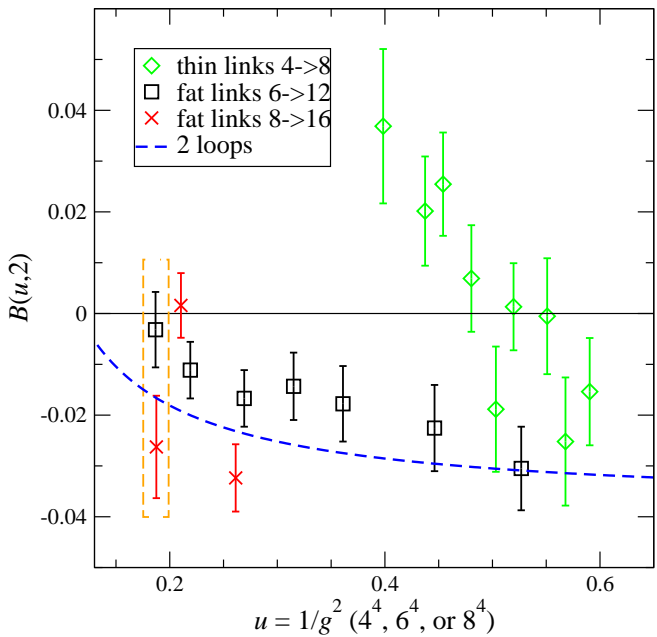}
\end{center}
\caption{Discrete beta function (\ref{eq:DBF}) for scale factor $s=2$. Thin link data are from Ref.~\cite{Shamir:2008pb}.
The bracketed points at left were measured in the metastable state at
$\beta=4.3$.
\label{fig:DBF}}
\end{figure}

In order to connect to our earlier work~\cite{Shamir:2008pb} on the thin-link lattice theory (on smaller lattices) we also plot the discrete beta function (DBF) as defined there,%
\footnote{The definition in Ref.~\cite{Shamir:2008pb} included
a factor of $K$ [see \Eq{defg} above] which we drop here.}
\bee
B(u,s)=\frac 1{g^2(sL)}-u,
\label{eq:DBF}
\ee
where $u=1/g^2(L)$ (Fig.~\ref{fig:DBF}.)
%%%%%%%%%%%%%%%%%%%%%%%%%%%%%%%%%%%%%%%%%%%%%%%%%%%%%%%%%%%%%%%%%%%%%
%%%%%%%%%%%%%%%%%%%%%%%%%%%%%%%%%%%%%%%%%%%%%%%%%%%%%%%%%%%%%%%%%%%%%
Again we note that the fat-link results show running that is slower than  two-loop perturbation theory, but no solid evidence for an IRFP.
In fact they rule out the fixed point found in the thin-link theory (on smaller lattices) at $1/g^2\simeq0.5$.

\begin{figure}
\begin{center}
\includegraphics[width=\columnwidth,clip]{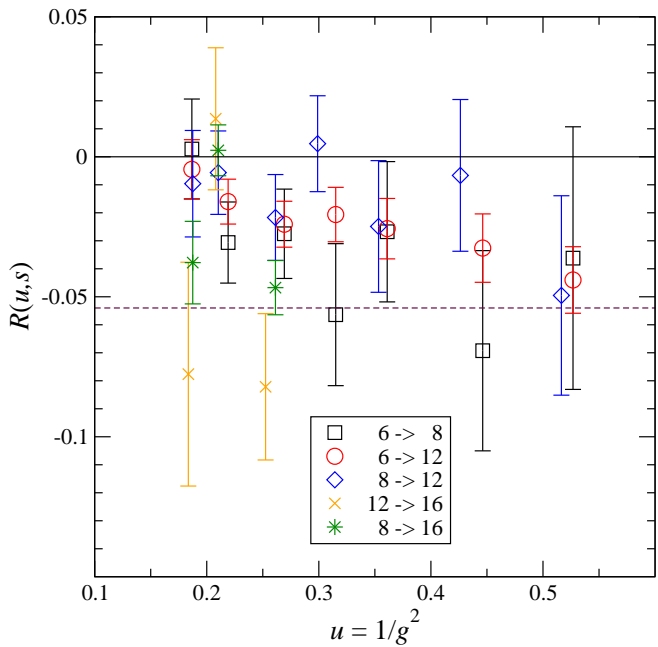}
\end{center}
\caption{Lattice approximants (\ref{eq:R}) to the beta function for many scale factors, derived by comparing lattices as shown in the legend.
The one-loop perturbative result $-13/(24\pi^2)\simeq-0.054$ is plotted as a dashed line.
\label{fig:effbeta}}
\end{figure}

The apparent linearity of the data in Fig.~\ref{fig:1g2vsl} suggests trying to collapse all the DBF's onto each other by plotting the ratio
\bee
R(u,s)\equiv\frac{B(u,s)}{\log s},
\label{eq:R}
\ee
combining data for all available scale factors $s$ on one graph.
$R(u,s)$ gives the beta function for $1/g^2$ when $s\to1$ in the continuum limit $L/a\to\infty$.
This is shown in Fig.~\ref{fig:effbeta}.
%%%%%%%%%%%%%%%%%%%%%%%%%%%%%%%%%%%%%%%%%%%%%%%%%%%%%%%%%%%%%%%%%%%%%
%%%%%%%%%%%%%%%%%%%%%%%%%%%%%%%%%%%%%%%%%%%%%%%%%%%%%%%%%%%%%%%%%%%%%
While, again, the data show slower running than the two-loop prediction, the scatter and the error bars pretty much preclude any ambitious further analysis.

%We are quite aware that the results of this analysis remain ambiguous with regard to the existence of a fixed point. 
A complete analysis of the data would involve taking the 
$a/L \rightarrow 0$ limit.
In principle, doing this on the weak coupling side of the phase transition would produce a result that is free of lattice discretization errors.
Fig.~\ref{fig:effbeta} shows that the scatter in $R(u,s)$ from different values of $a/L$ is at least as large
as the statistical fluctuations in the individual points. A fit to
$R(u,s,L)=R_0(u,s) + C/L$, for example, would just produce noise.
We are defeated by the slow intrinsic running of the coupling.
Nonetheless, we can at least make two plain statements:
\begin{itemize}
\item The IRFP observed in Ref.~\cite{Shamir:2008pb} is ruled out.
\item The SF coupling runs slowly over its observed range.
This slow running permits an easy and unambiguous measurement of the mass anomalous dimension as a function of the bare parameters or, equivalently, of the SF coupling $g^2$.
This is the subject of the next section.
\end{itemize}
%%%%%%%%%%%%%%%%%%%%%%%%%%%%%%%%%%%%%%%%%%%%%%%%%%%%%%%%%%%%%%%%%%%%%
\section{Mass anomalous dimension \label{sec:zp}}
%%%%%%%%%%%%%%%%%%%%%%%%%%%%%%%%%%%%%%%%%%%%%%%%%%%%%%%%%%%%%%%%%%%%%

After the discussion of the running gauge coupling, our result for the mass anomalous dimension is more definite:
$\gamma_m$ is never larger in magnitude than about~0.6.
This confirms the previous, noisy results of Ref.~\cite{DeGrand:2009hu}.
It suggests that, regardless of the existence of a zero of the beta function, this theory may not furnish a phenomenologically interesting model of walking technicolor.

We extract the anomalous dimension of $\bar\psi\psi$ from the scaling of $Z_P$ [\Eq{eq:ZP}] between systems rescaled as $L\to sL$.
We define the (continuum) mass step scaling function~\cite{Sint:1998iq,Capitani:1998mq,DellaMorte:2005kg,Bursa:2009we} as
\bee
  \label{eq:sigma_p}
  \sigma_P(u,s) = \left. {\frac{Z_P(sL)}{Z_P(L)}}
  \right|_{g^2(L)=u}.
\ee
It is related to the mass anomalous
dimension via
\bee
  \label{eq:sigPgamma}
  \sigma_P(u,s) = \exp\left[-\int_1^s \frac {dt}{t}\,\gamma_m\left(g^2(tL)\right)\right] .
\ee
Equation~(\ref{eq:sigPgamma}) is actually too complicated for our needs.
For any bare coupling $\beta$, the SF coupling $g^2(L)$ runs so slowly that we can replace \Eq{eq:sigPgamma} by
\bee
\sigma_P(u,s) = s^{ - \gamma_m(g^2)}.
  \label{eq:gamma}
\ee

\begin{figure}
\begin{center}
\includegraphics[width=\columnwidth,clip]{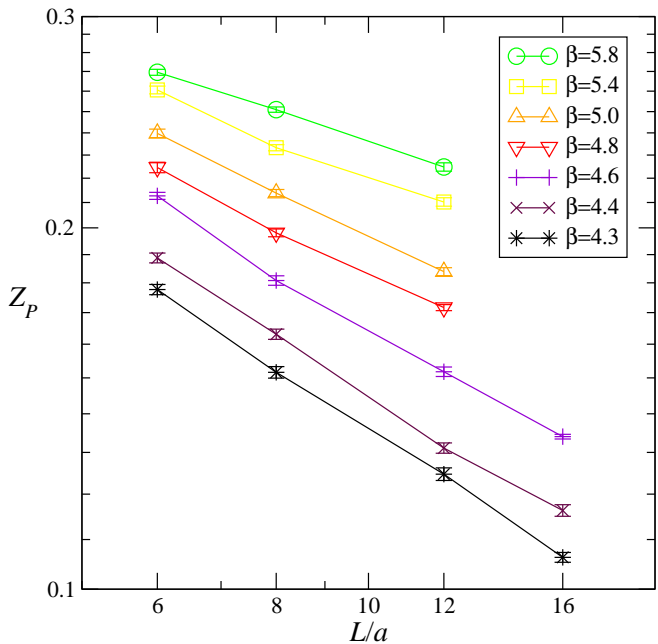}
\end{center}
\caption{Pseudoscalar renormalization constant $Z_P$ vs.~$L/a$.
\label{fig:zpl}}
\end{figure}

\begin{figure}
\begin{center}
\includegraphics[width=\columnwidth,clip]{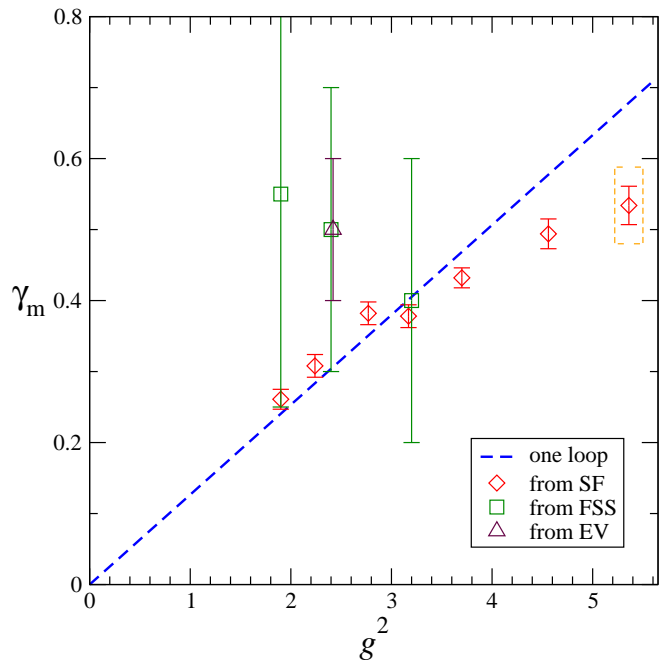}
\end{center}
\caption{Mass anomalous dimension $\gamma_m$ vs.~$g^2$, the $L=6a$ SF coupling.
The diamonds are the SF result (this paper).
The squares with the large error bars are the values of $\gamma_m$ derived
from the finite size scaling study of Ref.~{\protect{\cite{DeGrand:2009hu}}}
(plus one additional point added later), and the triangle
is the exponent inferred from the scaling of Dirac eigenvalues in that work.
The line is the perturbative prediction  $\gamma_m=6 C_2(R) g^2/(16\pi^2)$.
\label{fig:gmvsg2b}}
\end{figure}

Our results for $Z_P(L)$ are listed in Table~\ref{tab:zptable} and displayed in Fig.~\ref{fig:zpl}.
%%%%%%%%%%%%%%%%%%%%%%%%%%%%%%%%%%%%%%%%%%%%%%%%%%%%%%%%%%%%%%%%%%%%%
\begin{table}
\caption{Pseudoscalar renormalization factor $Z_P$ evaluated at the couplings $(\beta,\kappa_c)$, for the lattice sizes $L$ used in this study.}
\begin{center}
\begin{ruledtabular}
\begin{tabular}{cllll}
$\beta$ %& $\kappa$ 
&\multicolumn{4}{c}{$Z_P$}\\
\cline{2-5}
&   $L=6a$   & $L=8a$     & $L=12a$    & $L=16a$  \\
\hline
5.8 %& 0.12835 
& 0.2696(16) & 0.2509(12) & 0.2248(18) & \hfil-- \\
5.4 %& 0.12920 
& 0.2606(19) & 0.2333(14) & 0.2102(17) & \hfil-- \\
5.0 %& 0.13062 
& 0.2398(19) & 0.2318(15) & 0.1839(14) & \hfil-- \\
4.8 %& 0.13173 
& 0.2246(23) & 0.1981(15) & 0.1716(10) & \hfil-- \\
4.6 %& 0.13320 
& 0.2127(14) & 0.1808(16) & 0.1518(14) & 0.1340(6) \\
4.4 %& 0.13510 
& 0.1888(18) & 0.1631(16) & 0.1311(13) & 0.1163(13)  \\
4.3 %& 0.13617 
& 0.1777(17) & 0.1516(17) & 0.1247(15) & 0.1063(10) \\
\end{tabular}
\end{ruledtabular}
\end{center}
\label{tab:zptable}
\end{table}
%%%%%%%%%%%%%%%%%%%%%%%%%%%%%%%%%%%%%%%%%%%%%%%%%%%%%%%%%%%%%%%%%%%%%
%%%%%%%%%%%%%%%%%%%%%%%%%%%%%%%%%%%%%%%%%%%%%%%%%%%%%%%%%%%%%%%%%%%%%
%%%%%%%%%%%%%%%%%%%%%%%%%%%%%%%%%%%%%%%%%%%%%%%%%%%%%%%%%%%%%%%%%%%%%
As can be seen in the figure, the $L$-dependence of $Z_P$ is very close to linear on a log--log plot at all values of $\beta$.
This is a consequence of the slow running of the coupling constant.
The theory is ``conformal for all practical purposes'' over the range of $L$'s that are accessible at any single value of $\beta$. 
The slopes of the curves allow us to read off $\gamma_m(g^2)$ since the simplified \Eq{eq:gamma} implies
\bee
\log Z_P(L)=-\gamma_m \log L +\text{const}.
\ee
We thus fit straight lines to the data in Fig.~\ref{fig:zpl}, keeping only the three largest volumes at each $\beta$.
The result of the analysis is shown in Fig.~\ref{fig:gmvsg2b}, a plot of $\gamma_m(g^2)$ versus the SF coupling $g^2$.
We use $g^2(L=6a)$ for each bare coupling $\beta$; since $g^2(L)$ varies so little, this gives a good first approximation.
We compare to the one-loop perturbative expectation, $\gamma_m= 6 C_2(R) g^2/(16\pi^2)$.
It can be seen that the numerical results follow the perturbative line closely until they deviate downward at the strongest couplings.
The agreement with Ref.~\cite{DeGrand:2009hu} is gratifying, as is the much smaller uncertainty obtained with the present method.

%%%%%%%%%%%%%%%%%%%%%%%%%%%%%%%%%%%%%%%%%%%%%%%%%%%%%%%%%%%%%%%%%%%%%
%%%%%%%%%%%%%%%%%%%%%%%%%%%%%%%%%%%%%%%%%%%%%%%%%%%%%%%%%%%%%%%%%%%%%

\begin{figure}
\begin{center}
\includegraphics[width=0.9\columnwidth,clip]{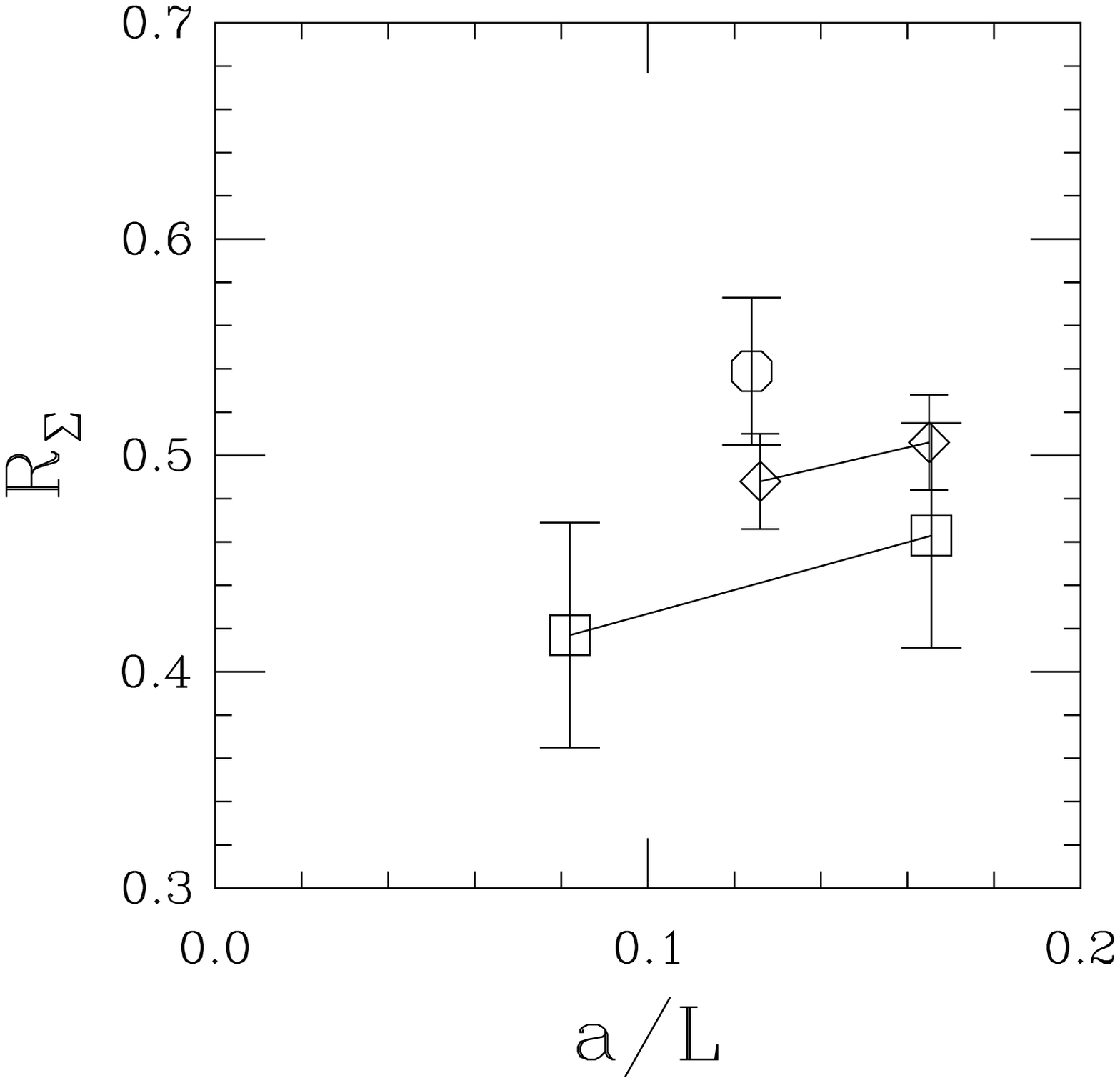}\\[10pt]
\includegraphics[width=0.9\columnwidth,clip]{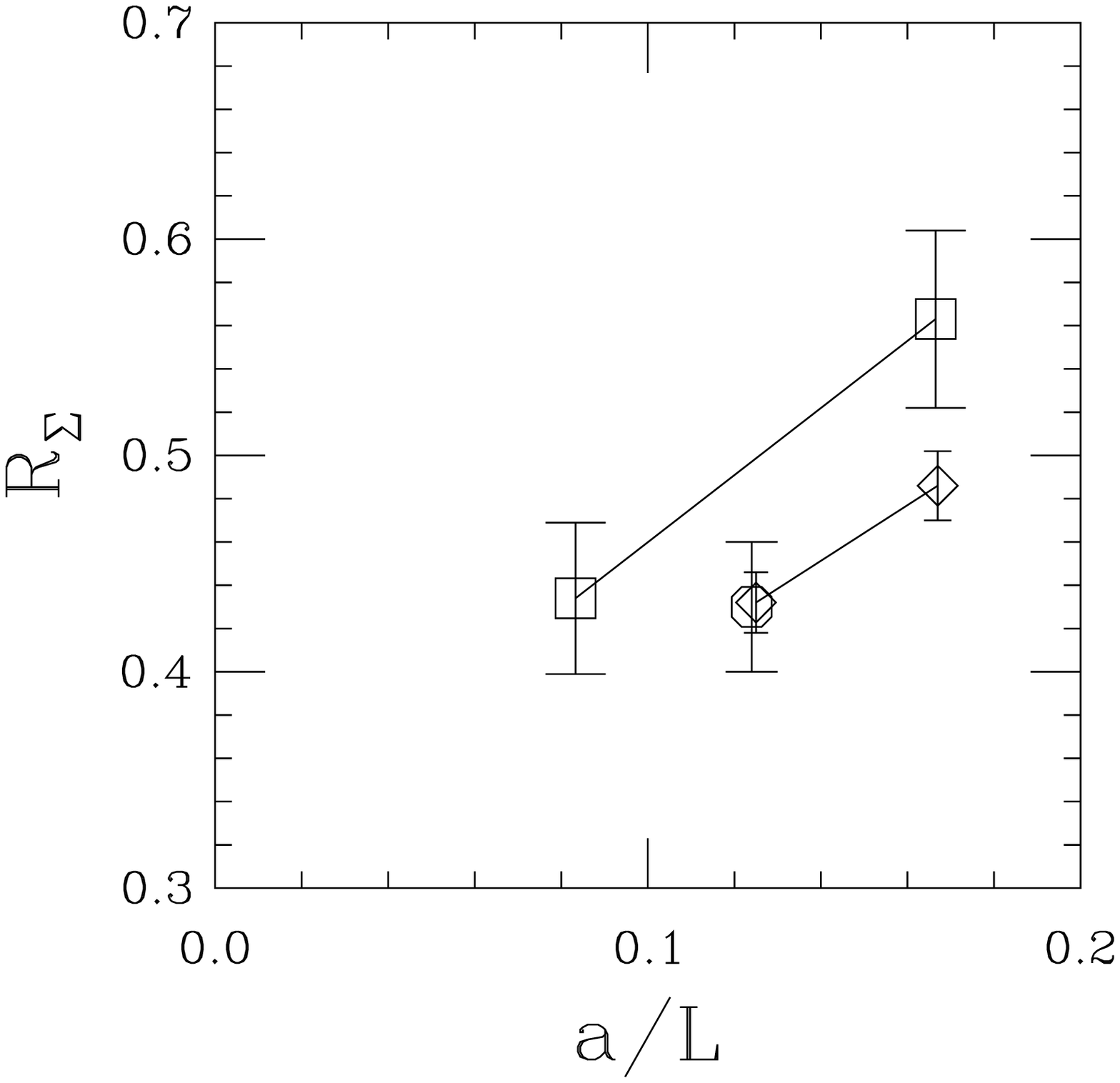}
\end{center}
\caption{Approximants $R_\Sigma(u,s,a/L)$ to the anomalous dimension $\gamma_m$ for the two bare couplings $\beta=4.4$ (top) and $\beta=4.6$ (bottom).
Diamonds are for $s=2$, squares for $s=4/3$ and the octagon is the $s=3/2$ point.
\label{fig:lscalss}}
\end{figure}

\begin{figure}
\begin{center}
\includegraphics[width=\columnwidth,clip]{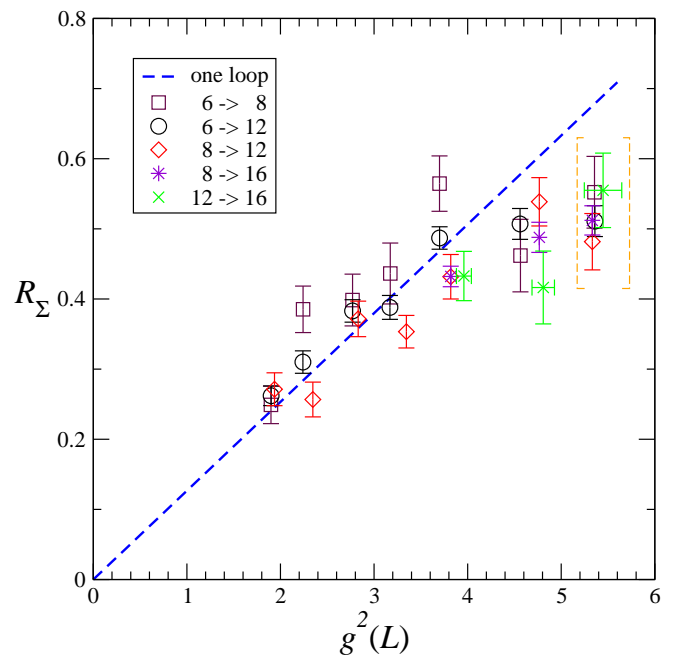}
\end{center}
\caption{Approximants $R_\Sigma(u,s,a/L)$ to the anomalous dimension $\gamma_m$, plotted against $u=g^2(L)$. In this cluttered graph, squares label points ($L/a=6$,$sL/a=8$), octagons (6,12), diamonds (8,12), bursts (8,16), and crosses, (12,16). The dashed line is the perturbative prediction.
\label{fig:effgamma}}
\end{figure}

To study finite-lattice effects we define the lattice approximation to the step scaling function (\ref{eq:sigma_p}),
\bee
  \Sigma_P(u,s,a/L)=\left. {\frac{Z_P(\beta,sL/a)}{Z_P(\beta,L/a)}}
  \right|_{{g}^2(L,a)=u},
  \label{eq:sigmaPdef}
\ee
such that 
\bee
  \label{eq:sigma_p2}
  \sigma_P(u,s) = \lim_{a\to 0}\Sigma_P(u,s,a/L).
\ee
We form the ratios of \Eq{eq:sigmaPdef} from the data in Table~\ref{tab:zptable}.
The rescaled quantities
\bee
R_\Sigma(u,s,a/L)=-\frac{\log\Sigma_P(u,s,a/L)}{\log s}
\ee
give $\gamma_m$ directly as $s\to1$ in the continuum limit.
They are shown for two bare couplings, $\beta=4.4$ and~4.6, in Fig.~\ref{fig:lscalss}.
At these couplings, we have data for $L/a=6$, 8, 12, and 16 and so we can form two combinations each with $s=2$ and $s=4/3$.
We also plot the analogous result for the $s=3/2$ pair $(L=8a,sL=12a)$.
While there does appear to be some cutoff ($a/L$) dependence, no possible extrapolation to $a/L=0$ can push $\gamma_m$ much above 0.6.
We follow up this figure with a compilation of $R_\Sigma(u,s,a/L)$, plotted against $u=g^2(L)$ for many pairs of $(L,sL)$, in Fig.~\ref{fig:effgamma}.

%%%%%%%%%%%%%%%%%%%%%%%%%%%%%%%%%%%%%%%%%%%%%%%%%%%%%%%%%%%%%%%%%%%%%
%%%%%%%%%%%%%%%%%%%%%%%%%%%%%%%%%%%%%%%%%%%%%%%%%%%%%%%%%%%%%%%%%%%%%
%%%%%%%%%%%%%%%%%%%%%%%%%%%%%%%%%%%%%%%%%%%%%%%%%%%%%%%%%%%%%%%%%%%%%

\begin{figure}
\begin{center}
\includegraphics[width=\columnwidth,clip]{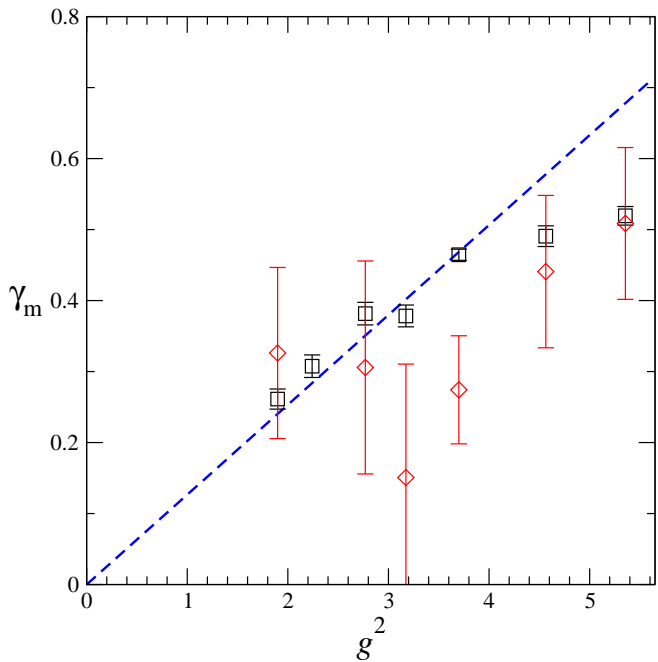}
\end{center}
\caption{$\gamma_m$ from fits to \Eq{eq:fitz}.
Squares are two-parameter fits to all data points at one $\beta$ value while diamonds are three-parameter fits.
The abscissa is the SF coupling determined for $L=6a$.
\label{fig:effgamma23}}
\end{figure}

We can use {\em all\/} the data to derive $\gamma_m(g^2)$ by fitting
at each $\beta$ to the form
\bee
\log Z_P= A(\beta)  - \gamma_m(\beta)\log L/a + C(\beta)\frac{a}{L}.
\label{eq:fitz}
\ee
The fit involves three parameters if $C$ is kept, two if it is discarded. 
Again, at any particular value of the bare coupling, the spread of $g^2(L)$ is so small as to be irrelevant in any plots.
Table~\ref{tab:gmtable} and Fig.~\ref{fig:effgamma23} show the results of two-parameter $(A,\gamma_m)$ and three parameter $(A,\gamma_m,C)$ fits to all (three or four) data points at each $\beta$ value.
Fitting all the data at one $\beta$ value to the three-parameter form produces results that are basically identical to those obtained if values of $R_\Sigma(u,s,a/L)$ are themselves used to approximate $\gamma_m(g^2,a)$ as above, whereupon $\gamma_m(g^2,a)$ is then fit to $\gamma_m(g^2,0) + C(a/L)$.
The data do not really allow us to say much, especially when we only have three $L$'s per $\beta$ value; nonetheless the figure does confirm that $\gamma_m$ never rises above~0.6.
Lattice spacing artifacts will not alter our result.

%%%%%%%%%%%%%%%%%%%%%%%%%%%%%%%%%%%%%%%%%%%%%%%%%%%%%%%%%%%%%%%%%%%%%
\begin{table}
\caption{The anomalous dimension $\gamma_m$ from two-parameter fits to the data
 in Table {\protect{\ref{tab:zptable}}}, showing the fit range and $\chi^2$.}
\begin{center}
\begin{ruledtabular}
\begin{tabular}{ccll}
$\beta$ %& $\kappa$ 
&$L$'s kept &$\gamma_m$ & $\chi^2$\\
\hline
5.8 %& 0.12835 
& 8, 12        & 0.271(23) & 0 \\
    %&           
& 6, 8, 12     & 0.261(14) & 0.29\\
5.4 %& 0.12920 
& 8, 12        & 0.257(25) & 0 \\
    %&           
& 6, 8, 12     & 0.308(15) & 7.12 \\
5.0 %& 0.13062  
& 8, 12        & 0.372(25) & 0 \\
    %&           
& 6, 8, 12     & 0.382(16) & 0.26 \\
4.8 %& 0.13173 
& 8, 12        & 0.353(23) & 0 \\
    %&           
& 6, 8, 12     & 0.378(15) & 2.04 \\
4.6 %& 0.13320 
& 6, 8, 12     & 0.432(14) & 0.0003\\
    %&           
& 6, 8, 12, 16 & 0.464(8)  &7.3 \\
4.4 %& 0.13510 
& 6, 8, 12     & 0.494(21) & 2.7 \\
    %&           
& 6, 8, 12, 16 & 0.491(14) & 2.7 \\
4.3 %& 0.13617 
& 6, 8, 12     & 0.512(20) & 0.77 \\
    %&           
& 6, 8, 12, 16 & 0.519(13) & 0.98 \\
\end{tabular}
\end{ruledtabular}
\end{center}
\label{tab:gmtable}
\end{table}
%%%%%%%%%%%%%%%%%%%%%%%%%%%%%%%%%%%%%%%%%%%%%%%%%%%%%%%%%%%%%%%%%%%%%
%%%%%%%%%%%%%%%%%%%%%%%%%%%%%%%%%%%%%%%%%%%%%%%%%%%%%%%%%%%%%%%%%%%%%
%%%%%%%%%%%%%%%%%%%%%%%%%%%%%%%%%%%%%%%%%%%%%%%%%%%%%%%%%%%%%%%%%%%%%

%%%%%%%%%%%%%%%%%%%%%%%%%%%%%%%%%%%%%%%%%%%%%%%%%%%%%%%%%%%%%%%%%%%%%
\section{Summary and conclusions  \label{sec:last}}
%%%%%%%%%%%%%%%%%%%%%%%%%%%%%%%%%%%%%%%%%%%%%%%%%%%%%%%%%%%%%%%%%%%%%

Our system exhibits two phases, a weak-coupling phase which is nonconfining and chirally restored, and a strong-coupling phase which is confining.
The AWI quark mass does not vanish in the confining phase, so we cannot perform a Schr\"odinger functional study there (except in a metastable state).
In the weak coupling phase, we find that the SF coupling runs more slowly than predicted by two-loop perturbation theory over the entire domain of couplings accessible to analysis.

The location of the fixed point reported in our earlier paper is not confirmed.
The present simulation has an improved lattice discretization and larger simulation volumes.
We did not observe an IRFP, nor did we rule one out.
We did not observe a positive beta function anywhere.

We studied the first-order phase boundary in some detail.
In order to ascertain the relevance of this transition
to continuum physics one has to study its location and strength
on yet larger lattices.
Moreover, the location of this transition almost certainly depends on the particular choice of bare action.
An obvious question is whether one can devise lattice discretizations that push the transition to larger values of the SF coupling.
Then it is possible that the extension of the beta function to larger couplings will reveal a fixed point.

Of other gauge theories similar to ours, the one with the most similar result is the SU(2) gauge theory with $N_f=2$ fermions in the adjoint representation---a candidate for ``minimal walking technicolor'' (MWT)~\cite{Bursa:2009we,DelDebbio:2010hu,DelDebbio:2010hx,Catterall:2007yx,Catterall:2008qk,Hietanen:2008mr,Hietanen:2009az,Catterall:2009sb}.
Its SF gauge coupling runs slowly, and evidence for a zero is ambiguous.
It has a small, nearly perturbative mass anomalous dimension. 
In these studies also Wilson fermions were used. 
There is a first-order transition as $\kappa$ is varied in strong coupling, but, in contrast to our results, this line ends in a critical point~\cite{Catterall:2008qk,Hietanen:2008mr}.
The connection of what is evidently a bulk transition to the finite-temperature phase transition is unclear.
Nonetheless, the critical point itself is a sign of interesting structure along the $\kappa_c$ line.
The latter is bounded by IR-repulsive fixed points at zero coupling (asymptotic freedom) and at strong coupling (the critical point), which may indicate that it is in the basin of attraction of an IR-attractive fixed point and hence that it constitutes a conformal phase.
We emphasize that, in contrast, our first-order transition at $\beta<\beta_1$ continues  smoothly into the first-order finite-temperature transition at $\kappa<\kappa_1$ without any sign of a critical point at the intersection with the $\kappa_c$ line.

The other class of slowly running theories is SU(3) gauge theory with $N_f>3$ flavors of light fermions in the {\em fundamental\/} representation.
The earliest study of the phase diagram using Wilson fermions is that of Iwasaki {\em et.~al\/} \cite{Iwasaki:1991mr,Iwasaki:2003de}. 
The authors found, for $7\le N_f\le 16$, that ``the massless quark line exists only in the deconfined phase.''
Recent simulations~\cite{Nagai:2009ip} in strong coupling for SU(2) and SU(3) gauge groups confirm this picture, with first order behavior (i.e., no $\kappa_c$ where $m_q=0$) appearing at $N_f=6$ for SU(2) and in the range 6--8 for SU(3).
Simulations with $N_f=10$ Wilson fermion flavors \cite{Yamada:2010wd} also produce a first order transition in strong coupling.

Kogut and Sinclair~\cite{Kogut:2010cz} have studied the SU(3) gauge theory with sextet fermions and reached conclusions rather different from ours. 
They report the existence of two finite-temperature transitions, a chiral and a confinement transition, at very different bare gauge couplings, implying both dramatic scale separation and non-conformal physics.
That study was done with unimproved staggered fermions and the usual square-root prescription.
Away from the continuum limit, this prescription is known to induce non-localities;
equivalently, the number of quark species is not well defined.%
\footnote{For a recent review, see Ref.~\cite{Golterman:2008gt}.}
It is important to monitor the magnitude of taste violation carefully in such calculations.
Badly broken taste symmetry would mean that the theory under study has fewer effective massless flavors, which would bias the result towards confinement rather than conformality.

Returning to the present study, we note again that the slow running of the gauge coupling constant allows an inexpensive and  accurate measurement of the mass anomalous dimension $\gamma_m$.
This measurement confirms, and considerably improves on, the calculation of Ref.~\cite{DeGrand:2009hu}.
$\gamma_m$ is small, never greater than~0.6 in the coupling region where we can measure it.

%{\em What the hell is this paragraph about?  It doesn't describe Bursa or the other Del Debbio papers at all!}
%The recent determination \cite{Bursa:2009we,DelDebbio:2010hu,DelDebbio:2010hx} of the mass anomalous dimension in the MWT model (for one $\beta$ value) gave $\gamma_m=0.22(6)$. 
%This is a somewhat smaller value than what we have found here, and, again, much less than unity.

The small value of $\gamma_m$ in this theory presents a challenge for technicolor phenomenology.
Even if sextet QCD is not phenomenologically viable, however, it is characterized by a coupling constant that evolves very slowly with scale.
Such systems are theoretically interesting in their own right.

%%%%%%%%%%%%%%%%%%%%%%%%%%%%%%%%%%%%%%%%%%%%%%%%%%%%%%%%%%%%%%%%%%%%%
\begin{acknowledgments}
%%%%%%%%%%%%%%%%%%%%%%%%%%%%%%%%%%%%%%%%%%%%%%%%%%%%%%%%%%%%%%%%%%%%%%
We thank
S.~Chivukula,
P.~Damgaard,
A.~Hasenfratz,
D.~Kaplan,
A.~Nelson,
and
E.~Simmons
for discussions and correspondence.
T.~D. is grateful for the hospitality of the Niels Bohr International Academy during part of
the time he carried out this research.
B.~S. and Y.~S. thank the University of Colorado for hospitality.
This work was supported in part by the Israel Science Foundation
under grants no.~173/05 and no.~423/09 and by the US Department of Energy.  
This research was also supported in part by the National Science Foundation through TeraGrid resources provided by the University of Texas under grants no.~TG-PHY080042 and no.~TG-PHY090023.
Further computations were done on clusters at the University of Colorado and at Tel Aviv University. 

Our computer code is based on the publicly available package of the MILC collaboration~\cite{MILC}.
The code for hypercubic smearing was adapted from a program written by A.~Hasenfratz, R.~Hoffmann and S.~Schaefer~\cite{Hasenfratz:2008ce}.

%%%%%%%%%%%%%%%%%%%%%%%%%%%%%%%%%%%%%%%%%%%%%%%%%%%%%%%%%%%%%%%%%%%%%
\end{acknowledgments}
%%%%%%%%%%%%%%%%%%%%%%%%%%%%%%%%%%%%%%%%%%%%%%%%%%%%%%%%%%%%%%%%%%%%%
\appendix*
\section{Illustrations of the Strong Coupling Transition\label{appB}}
%%%%%%%%%%%%%%%%%%%%%%%%%%%%%%%%%%%%%%%%%%%%%%%%%%%%%%%%%%%%%%%%%%%%%
\begin{figure}
\begin{center}
\includegraphics[width=0.9\columnwidth,clip]{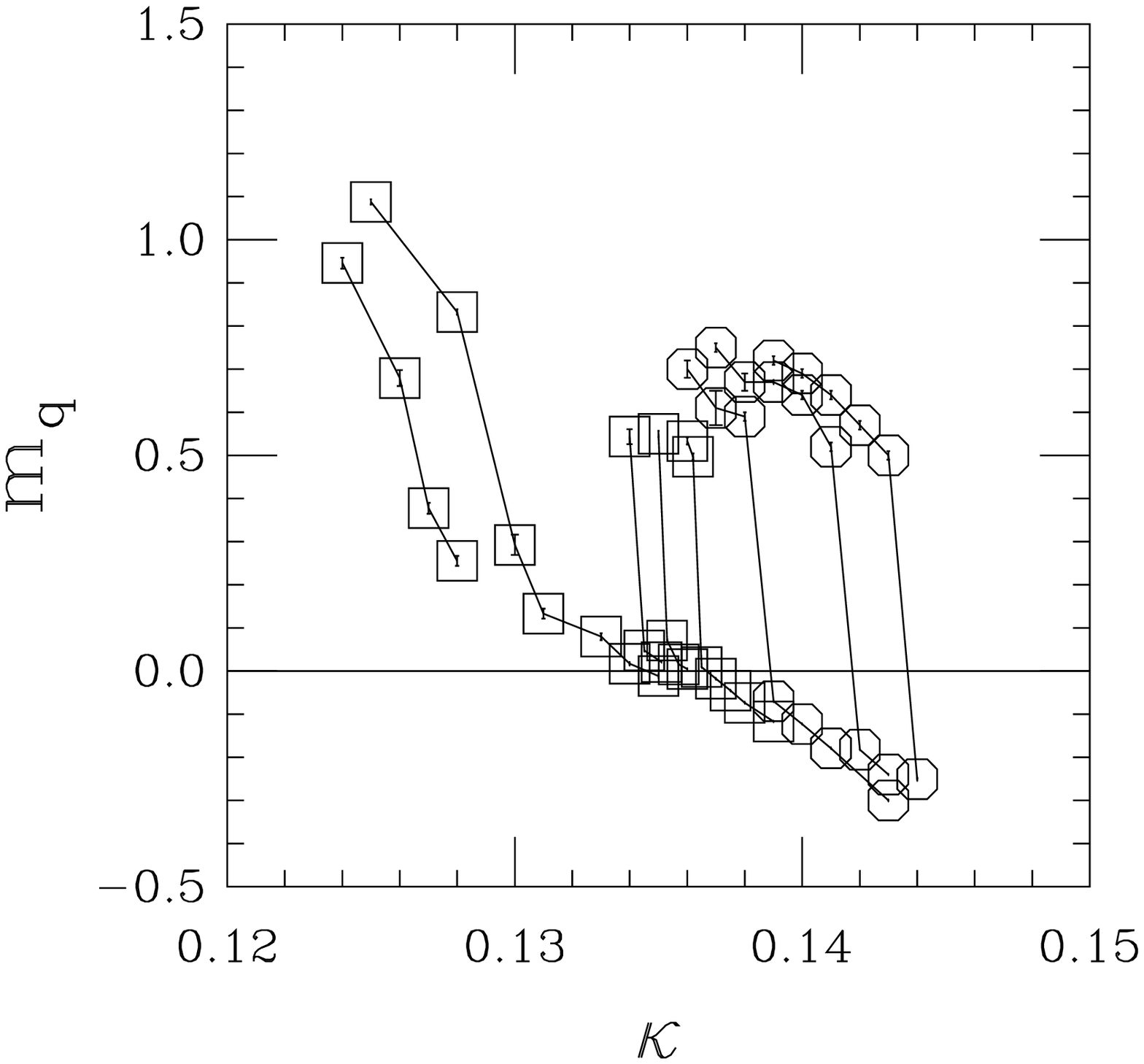}\\[10pt]
\includegraphics[width=0.9\columnwidth,clip]{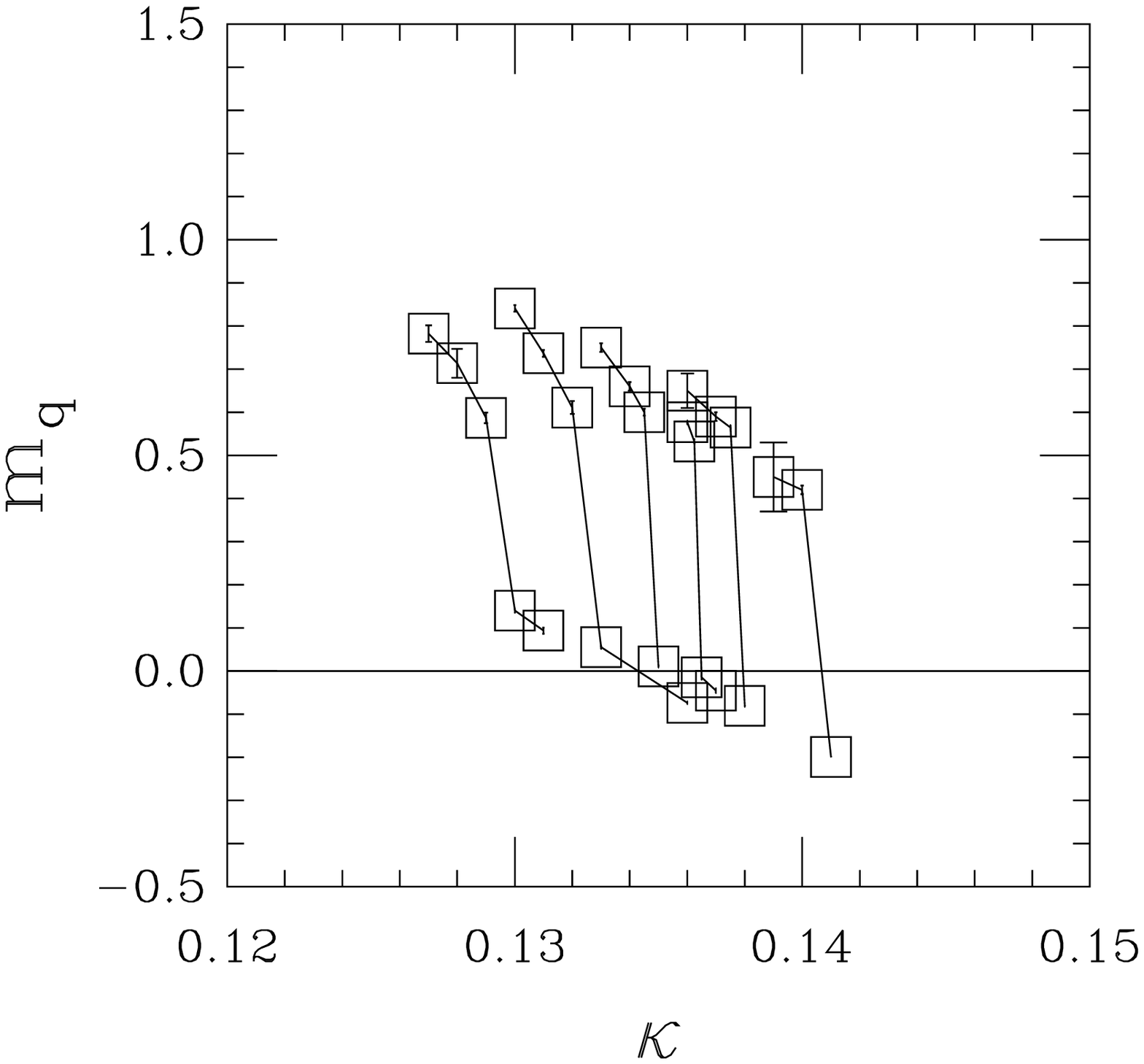}
\end{center}
\caption{The AWI quark mass $m_q$ from simulations on $N_t=6$ (top) and $N_t=8$ (bottom) lattices.
Data collected at the same $\beta$ values are connected by lines.
In the top panel, squares indicate $6^4$ volumes while octagons indicate $12\times 6^3$.
The $\beta$ values are, from left to right, $\beta=4.6$, 4.5, 4.4, 4.35, 4.3, 4.2, 4.1, and~4.0.
In the bottom panel, all data are from $8^4$ lattices and the $\beta$ values are, from left to right, $\beta=4.6$, 4.5, 4.4, 4.35, 4.3, and~4.2.
\label{fig:mqlowb}}
\end{figure}
%%%%%%%%%%%%%%%%%%%%%%%%%%%%%%%%%%%%%%%%%%%%%%%%%%%%%%%%%%%%%%%%%%%%%
In this appendix we present data supporting the location of the first-order phase transitions $\kappa_{\text{conf}}(\beta)$ for $N_t=6$ and~8, as well as determinations of where the boundaries cross the $\kappa_c(\beta)$ curve.
(We denote the crossing points by $[\beta_1(L),\kappa_1(L)]$, for $L=6a$ and $8a$.)

\begin{figure}[tb]
\begin{center}
\includegraphics[width=0.9\columnwidth,clip]{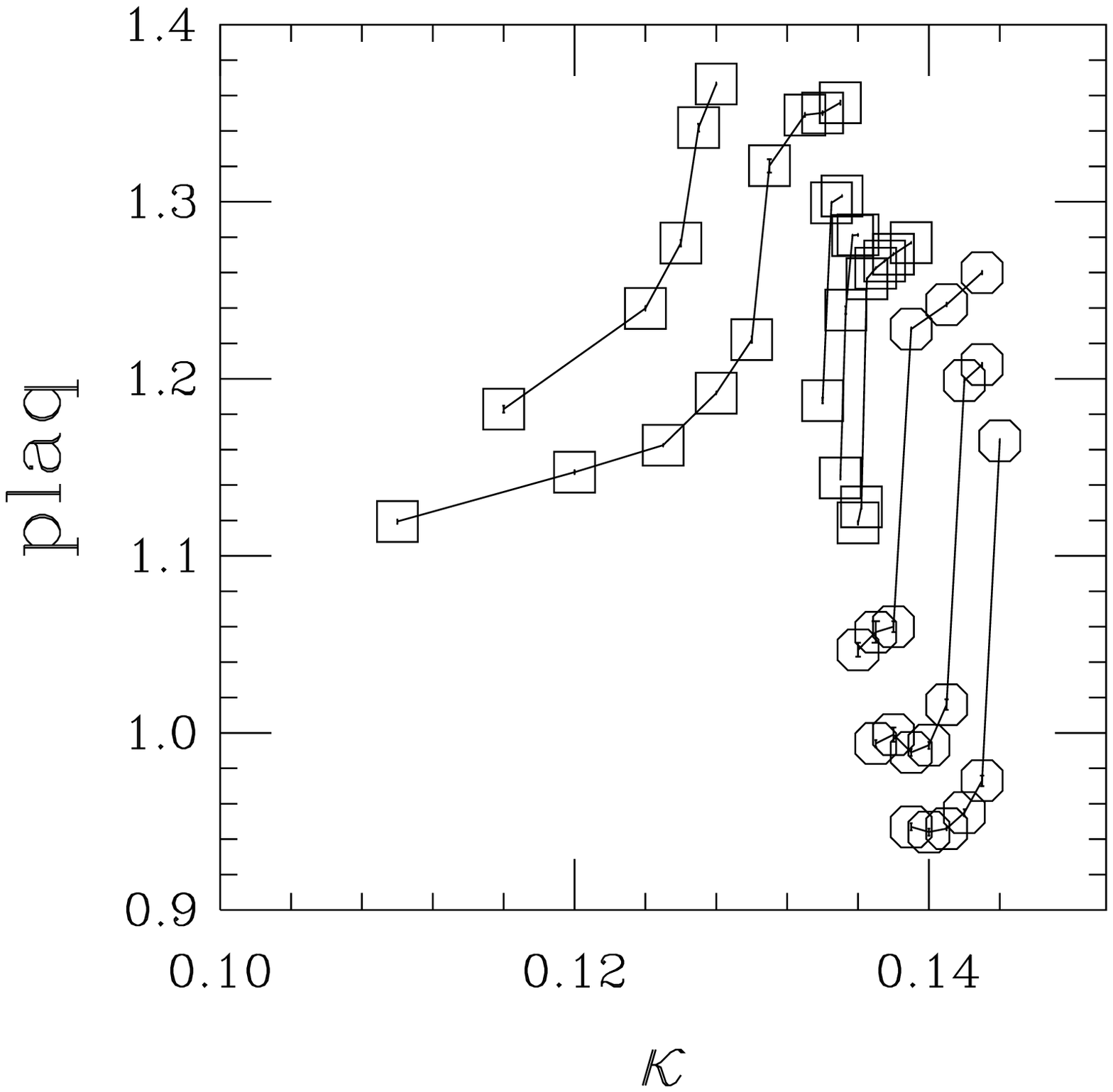}\\[10pt]
\includegraphics[width=0.9\columnwidth,clip]{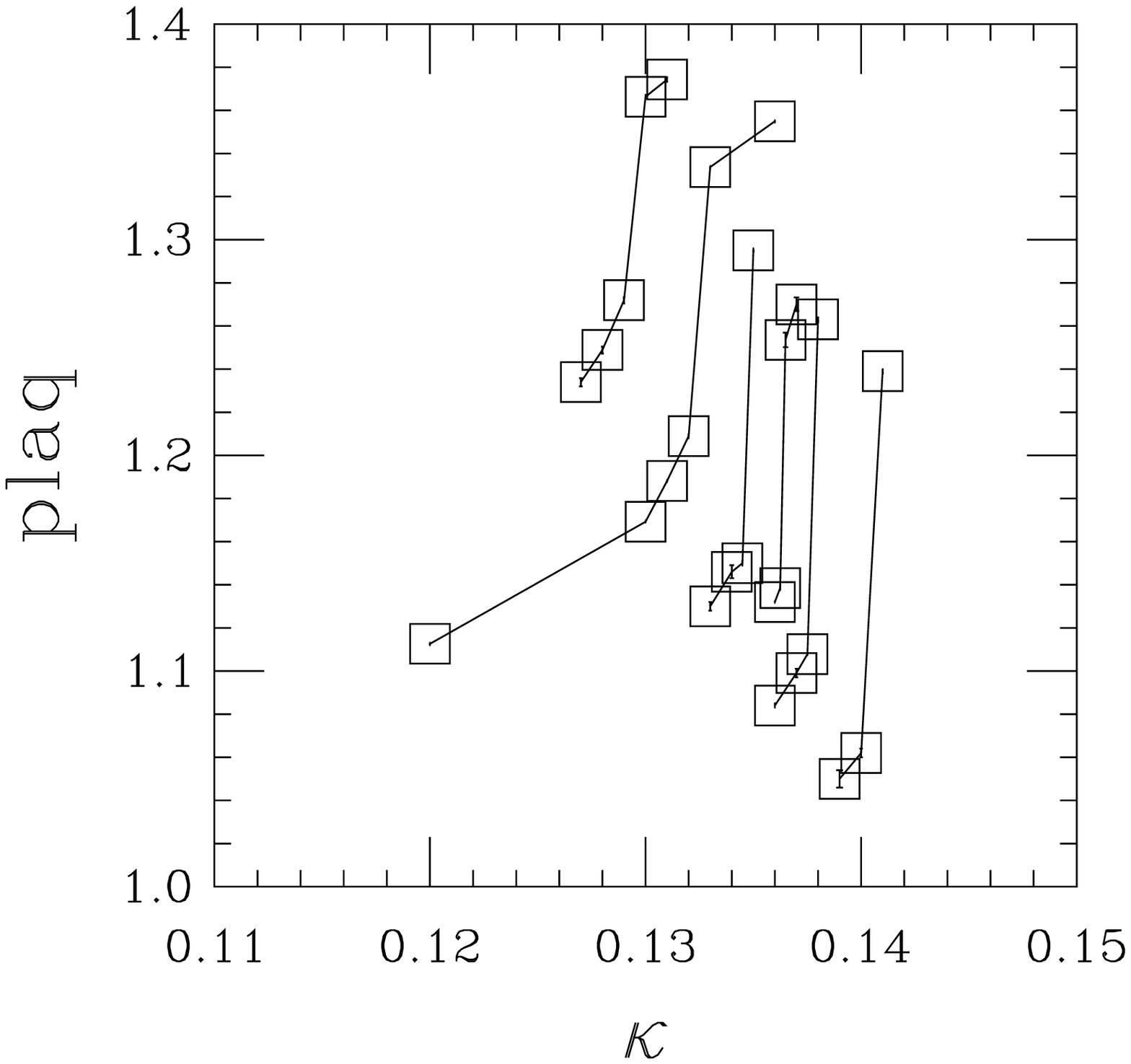}
\end{center}
\caption{The plaquette average from the same simulations as in Fig.~\ref{fig:mqlowb}. 
Symbols and $\beta$ values are the same.
Data collected at the same $\beta$ values are connected by lines.
\label{fig:plaqlowb}}
\end{figure}

First we show data for the AWI quark mass $m_q$ (Fig.~\ref{fig:mqlowb}) and plaquette average (Fig.~\ref{fig:plaqlowb})
from simulations with SF boundary conditions.
Each curve shows the variation with $\kappa$ for a given $\beta$, as $\kappa$ is swept across the phase boundary at $\kappa_{\text{conf}}$.
The first-order transition is evident in each curve, as is the fact that the discontinuity nowhere tends to zero:  There is no critical point anywhere.

%%%%%%%%%%%%%%%%%%%%%%%%%%%%%%%%%%%%%%%%%%%%%%%%%%%%%%%%%%%%%%%%%%%%%
%%%%%%%%%%%%%%%%%%%%%%%%%%%%%%%%%%%%%%%%%%%%%%%%%%%%%%%%%%%%%%%%%%%%%
In Fig.~\ref{fig:mqlowb} we see that the condition $m_q=0$ can be used to define $\kappa_c(\beta)$ for $\beta>\beta_1(L)$, where $4.30<\beta_1(6)<4.35$ while $4.35<\beta_1(8)<4.40$.  
Thus the phase boundary and the $\kappa_c$ curve are distinct for $\beta>\beta_1(L)$, as seen in Figs.~\ref{sketch} and~\ref{fig:bktSF}.
For $\beta<\beta_1(L)$ we find that $m_q$ crosses zero discontinuously, and thus there is no $\kappa_c$ curve here.
(The data shown here refer to equilibrium states only; the metastable extension of the $\kappa_c$ curve to the left of $\beta_1$ cannot be seen.)

%%%%%%%%%%%%%%%%%%%%%%%%%%%%%%%%%%%%%%%%%%%%%%%%%%%%%%%%%%%%%%%%%%%%%
\begin{figure}
\begin{center}
\includegraphics[width=0.9\columnwidth,clip]{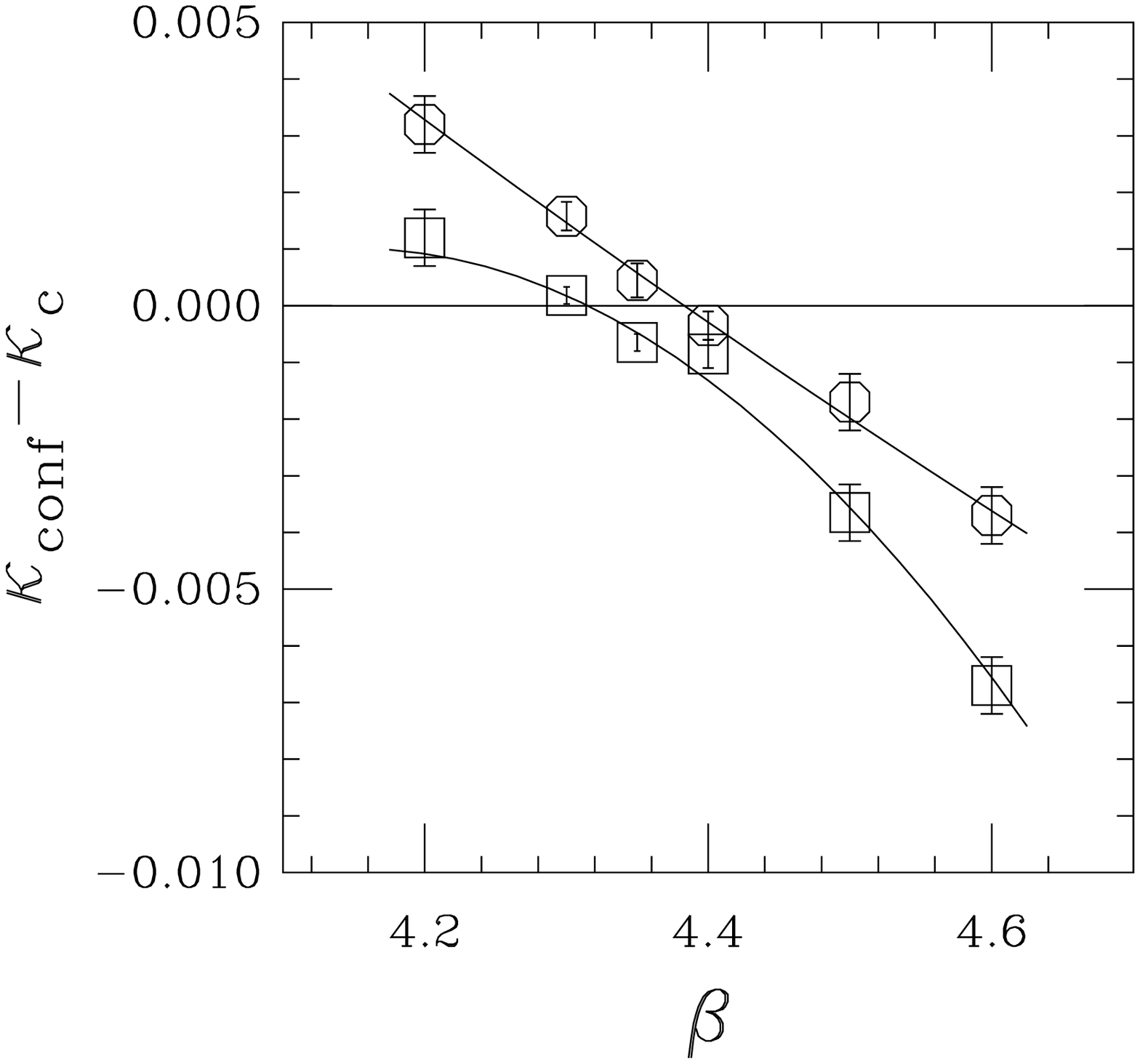}
\end{center}
\caption{Finding the intersection $(\beta_1,\kappa_1)$ of the first-order phase boundary $\kappa_{\text{conf}}(\beta)$ with $\kappa_c(\beta)$.
The data come from simulations with SFBC (cf.~Fig.~\ref{fig:bktSF}).
The curves are quadratic fits through the data.
Squares show $N_t=6$ data; octagons, $N_t=8$.
\label{fig:fitbetac}}
\end{figure}

Since the variation of $\beta_1(L)$ is of great importance, we devote some effort
 to interpolating $\kappa_{\text{conf}}(\beta)$ to find its intersection with the $\kappa_c$ curve.
In Fig.~\ref{fig:fitbetac} we plot the difference
 $\kappa_{\text{conf}}-\kappa_c$ vs.~$\beta$, where $\kappa_c$ is
 determined in the deconfined phase.
(Where this quantity is positive, the phase boundary lies above the
 $\kappa_c$ curve; the AWI quark mass in the deconfined phase at the
 phase boundary  is negative, and the $m_q=0$  point lies in a metastable phase.)
The curves in the figure are quadratic fits.
Inverting them to find $\beta_1$, the point where the transition crosses
 the $\kappa_c$ curve, we find $\beta_1=4.315(8)$ for $N_t=6$ (with $\chi^2=5.26/3$~dof)
and 4.383(10) for $N_t=8$ (with $\chi^2=0.83$).
The difference is then $\Delta\beta=0.068(13)$, to be compared to the prediction of 0.08 from \Eq{eq:twoloopshift}.
These results are stable under variation in the number of points used in the fit:
For example, quadratic fits to the three points at $\beta=4.3$, 4.35 and~4.4 give 4.308(6) and 4.376(15) for $N_t=6,8$, respectively.

%%%%%%%%%%%%%%%%%%%%%%%%%%%%%%%%%%%%%%%%%%%%%%%%%%%%%%%%%%%%%%%%%%%%%
%\begin{figure}
%\begin{center}
%\includegraphics[width=0.9\columnwidth,clip]{fit_betac3.eps}
%\end{center}
%\caption{
%Quadratic fits through the three points at $\beta=4.3$, 4.4, and~4.5.
%Squares show $N_t=6$ data; octagons, $N_t=8$.
%\label{fig:fitbetac3}}
%\end{figure}
%%%%%%%%%%%%%%%%%%%%%%%%%%%%%%%%%%%%%%%%%%%%%%%%%%%%%%%%%%%%%%%%%%%%%

%%%%%%%%%%%%%%%%%%%%%%%%%%%%%%%%%%%%%%%%%%%%%%%%%%%%%%%%%%%%%%%%%%%%%

%%%%%%%%%%%%%%%%%%%%%%%%%%%%%%%%%%%%%%%%%%%%%%%%%%%%%%%%%

\begin{thebibliography}{99}
%%%%%%%%%%%%%%%%%%%%%%%%%%%%%%%%%%%%%%%%%%%%%%%%%%%%%%%%%%%%%%%%%%%%%
  

%%%%%%%%%%%%%%%%%%%%%%%%%%%%%%%
% introduction refs
%%%%%%%%%%%%%%%%%%%%%%%%%%%%%%
%\cite{Hill:2002ap}
\bibitem{Hill:2002ap}
  C.~T.~Hill and E.~H.~Simmons,
  ``Strong dynamics and electroweak symmetry breaking,''
  Phys.\ Rept.\  {\bf 381}, 235 (2003)
  [Erratum-ibid.\  {\bf 390}, 553 (2004)]
  [arXiv:hep-ph/0203079];
  %%CITATION = PRPLC,381,235;%%

%\cite{Georgi:2007ek}
\bibitem{Georgi:2007ek}
  H.~Georgi,
  ``Unparticle Physics,''
  Phys.\ Rev.\ Lett.\  {\bf 98}, 221601 (2007)
  [arXiv:hep-ph/0703260];
  %%CITATION = PRLTA,98,221601;%%
  ``Another Odd Thing About Unparticle Physics,''
  Phys.\ Lett.\  B {\bf 650}, 275 (2007)
  [arXiv:0704.2457 [hep-ph]].
  %%CITATION = PHLTA,B650,275;%%

\bibitem{reviews}
For reviews see
  E.~Pallante,
  ``Strongly and slightly flavored gauge theories,''
  arXiv:0912.5188 [hep-lat];
  %%CITATION = ARXIV:0912.5188;%%
  G.~T.~Fleming,
  ``Strong Interactions for the LHC,''
  PoS {\bf LATTICE2008}, 021 (2008)
  [arXiv:0812.2035 [hep-lat]].
  %%CITATION = POSCI,LATTICE2008,021;%%

\bibitem{Caswell:1974gg}
  W.~E.~Caswell,
  ``Asymptotic behavior of nonabelian gauge theories to two loop order,''
  Phys.\ Rev.\ Lett.\  {\bf 33}, 244 (1974).
  %%CITATION = PRLTA,33,244;%%
  
\bibitem{Banks:1981nn}
  T.~Banks and A.~Zaks,
  ``On the phase structure of vector-like gauge theories with massless fermions,''
  Nucl.\ Phys.\  B {\bf 196}, 189 (1982).
  %%CITATION = NUPHA,B196,189;%%

%\cite{Holdom:1981rm}
\bibitem{Holdom:1981rm}
  B.~Holdom,
  ``Raising The Sideways Scale,''
  Phys.\ Rev.\  D {\bf 24}, 1441 (1981).
  %%CITATION = PHRVA,D24,1441;%%

%\cite{Yamawaki:1985zg}
\bibitem{Yamawaki:1985zg}
  K.~Yamawaki, M.~Bando and K.~i.~Matumoto,
  %``Scale Invariant Technicolor Model And A Technidilaton,''
  Phys.\ Rev.\ Lett.\  {\bf 56}, 1335 (1986).
  %%CITATION = PRLTA,56,1335;%%

%\cite{Appelquist:1997fp}
\bibitem{Appelquist:1997fp}
  T.~Appelquist, J.~Terning and L.~C.~R.~Wijewardhana,
  ``Postmodern technicolor,''
  Phys.\ Rev.\ Lett.\  {\bf 79}, 2767 (1997)
  [arXiv:hep-ph/9706238].
  %%CITATION = PRLTA,79,2767;%%

%\cite{Chivukula:2010tn}
\bibitem{Chivukula:2010tn}
  R.~S.~Chivukula and E.~H.~Simmons,
  ``Condensate enhancement and $D$-meson mixing in technicolor theories,''
  arXiv:1005.5727 [hep-lat].
  %%CITATION = ARXIV:1005.5727;%%

%\cite{Nelson:2006zz}
\bibitem{Nelson:2006zz}
  A.~E.~Nelson,
  ``Lattice calculations for physics beyond the standard model,''
  PoS {\bf LAT2006}, 016 (2006).
  %%CITATION = POSCI,LAT2006,016;%%

\bibitem{Sannino:2004qp}
F.~Sannino and K.~Tuominen,
  ``Orientifold theory dynamics and symmetry breaking,''
  Phys.\ Rev.\  D {\bf 71}, 051901 (2005)
  [arXiv:hep-ph/0405209].
  %%CITATION = PHRVA,D71,051901;%%

%\cite{Hong:2004td}
\bibitem{Hong:2004td}
  D.~K.~Hong, S.~D.~H.~Hsu and F.~Sannino,
  ``Composite Higgs from higher representations,''
  Phys.\ Lett.\  B {\bf 597}, 89 (2004)
  [arXiv:hep-ph/0406200].
  %%CITATION = PHLTA,B597,89;%%

\bibitem{Dietrich:2005jn}
D.~D.~Dietrich, F.~Sannino and K.~Tuominen,
  ``Light composite Higgs from higher representations versus electroweak precision measurements: Predictions for LHC,''
  Phys.\ Rev.\  D {\bf 72}, 055001 (2005)
  [arXiv:hep-ph/0505059].
  %%CITATION = PHRVA,D72,055001;%%

%\cite{Belyaev:2008yj}
\bibitem{Belyaev:2008yj}
  A.~Belyaev, R.~Foadi, M.~T.~Frandsen, M.~Jarvinen, F.~Sannino and A.~Pukhov,
  ``Technicolor walks at the LHC,''
  Phys.\ Rev.\  D {\bf 79}, 035006 (2009)
  [arXiv:0809.0793 [hep-ph]].
  %%CITATION = PHRVA,D79,035006;%%

%%%%%%%%%%%%%%%%%%%%%%%%% us %%%%%%%%%%%%%%%%%%%%%%%%%%%

%\cite{Shamir:2008pb}
\bibitem{Shamir:2008pb}
  Y.~Shamir, B.~Svetitsky and T.~DeGrand,
  ``Zero of the discrete beta function in SU(3) lattice gauge theory with color sextet fermions,''
  Phys.\ Rev.\  D {\bf 78}, 031502 (2008)
  [arXiv:0803.1707 [hep-lat]].
  %%CITATION = PHRVA,D78,031502;%%

%\cite{DeGrand:2008kx}
\bibitem{DeGrand:2008kx}
  T.~DeGrand, Y.~Shamir and B.~Svetitsky,
  ``Phase structure of SU(3) gauge theory with two flavors of symmetric-representation fermions,''
  Phys.\ Rev.\  D {\bf 79}, 034501 (2009)
  [arXiv:0812.1427 [hep-lat]].
  %%CITATION = PHRVA,D79,034501;%%

%\cite{DeGrand:2009hu}
\bibitem{DeGrand:2009hu}
  T.~DeGrand,
  ``Finite-size scaling tests for SU(3) lattice gauge theory with color sextet fermions,''
  Phys.\ Rev.\  D {\bf 80}, 114507 (2009)
  [arXiv:0910.3072 [hep-lat]].
  %%CITATION = PHRVA,D80,114507;%%

%%%%%%%%%%%%%%%%%%%%%%%%% more intro %%%%%%%%%%%%%%%%%%%%%%%%%%%

%\cite{Fodor:2009ar}
\bibitem{Fodor:2009ar}
  Z.~Fodor, K.~Holland, J.~Kuti, D.~N\'ogradi and C.~Schroeder,
  ``Chiral properties of SU(3) sextet fermions,''
  JHEP {\bf 0911}, 103 (2009)
  [arXiv:0908.2466 [hep-lat]].
  %%CITATION = JHEPA,0911,103;%%

%\cite{Machtey:2009wu}
\bibitem{Machtey:2009wu}
  O.~Machtey and B.~Svetitsky,
  ``Metastable nonconfining states in SU(3) lattice gauge theory with sextet fermions,''
  Phys.\ Rev.\  D {\bf 81}, 014501 (2010)
  [arXiv:0911.0886 [hep-lat]].
  %%CITATION = PHRVA,D81,014501;%%

%\cite{Kogut:2010cz}
\bibitem{Kogut:2010cz}
  J.~B.~Kogut and D.~K.~Sinclair,
  ``Thermodynamics of lattice QCD with 2 flavours of colour-sextet quarks: A model of walking/conformal Technicolor,''
  arXiv:1002.2988 [hep-lat].
  %%CITATION = ARXIV:1002.2988;%%

%\cite{Iwasaki:2003de}
\bibitem{Iwasaki:2003de}
  Y.~Iwasaki, K.~Kanaya, S.~Kaya, S.~Sakai and T.~Yoshi\'e,
  ``Phase structure of lattice QCD for general number of flavors,''
  Phys.\ Rev.\  D {\bf 69}, 014507 (2004)
  [arXiv:hep-lat/0309159].
  %%CITATION = PHRVA,D69,014507;%%

%%%%%%%%%%%%%%%%%%%%%%%%%%%%%%%%%%%%%%%%%%%%%%%%%%%%%%%%%
%method refs
%%%%%%%%%%%%%%%%%%%%%%%%%%%%%%%%%%%%%%%%%%%%%%%%%%%%%%%%%
\bibitem{Sheikholeslami:1985ij}
  B.~Sheikholeslami and R.~Wohlert,
  ``Improved continuum limit lattice action for QCD with Wilson fermions,''
  Nucl.\ Phys.\  B {\bf 259}, 572 (1985).
  %%CITATION = NUPHA,B259,572;%%

%\cite{Hasenfratz:2007rf}
\bibitem{Hasenfratz:2007rf}
  A.~Hasenfratz, R.~Hoffmann and S.~Schaefer,
  ``Hypercubic smeared links for dynamical fermions,''
  JHEP {\bf 0705}, 029 (2007)
  [arXiv:hep-lat/0702028].
%%CITATION = JHEPA,0705,029;%%

%\cite{Hasenbusch:2001ne}
\bibitem{Hasenbusch:2001ne}
  M.~Hasenbusch,
  ``Speeding up the Hybrid-Monte-Carlo algorithm for dynamical fermions,''
  Phys.\ Lett.\  B {\bf 519}, 177 (2001)
  [arXiv:hep-lat/0107019].
  %%CITATION = PHLTA,B519,177;%%

%\cite{Urbach:2005ji}
\bibitem{Urbach:2005ji}
 C.~Urbach, K.~Jansen, A.~Shindler and U.~Wenger,
 ``HMC algorithm with multiple time scale integration and mass preconditioning,''
 Comput.\ Phys.\ Commun.\  {\bf 174}, 87 (2006)
 [arXiv:hep-lat/0506011].
 %%CITATION = CPHCB,174,87;%%

%\cite{Takaishi:2005tz}
\bibitem{Takaishi:2005tz}
  T.~Takaishi and P.~de Forcrand,
  ``Testing and tuning new symplectic integrators for hybrid Monte Carlo algorithm in lattice QCD,''
  Phys.\ Rev.\  E {\bf 73}, 036706 (2006)
  [arXiv:hep-lat/0505020].
  %%CITATION = PHRVA,E73,036706;%%

%%%%%%%%%%%%%%%%%%%%%%%%%%%%%%%%%%%%%%%%%%%%%%%%%%%%%%%
%Fat link justification
%%%%%%%%%%%%%%%%%%%%%%%%%%%%%%%%%%%%%%%%%%%%%%%%%%%%%%%%%

%\cite{Hoffmann:2007nm}
\bibitem{Hoffmann:2007nm}
  R.~Hoffmann, A.~Hasenfratz and S.~Schaefer,
  ``Non-perturbative improvement of nHYP smeared Wilson fermions,''
  PoS {\bf LAT2007}, 104 (2007)
  [arXiv:0710.0471 [hep-lat]].
  %%CITATION = POSCI,LAT2007,104;%%

\bibitem{evgeny}
Y.~Shamir, B.~Svetitsky, and E.~Yurkovsky, in preparation.

%\cite{DeGrand:2002vu}
\bibitem{DeGrand:2002vu}
See the tables in
  T.~A.~DeGrand, A.~Hasenfratz and T.~G.~Kovacs,
  ``Improving the chiral properties of lattice fermions,''
  Phys.\ Rev.\  D {\bf 67}, 054501 (2003)
  [arXiv:hep-lat/0211006].
  %%CITATION = PHRVA,D67,054501;%%
For nonperturbative determinations (with fundamental representation fermions), see Ref.~\cite{Hasenfratz:2008ce}.

%%%%%%%%%%%%%%%%%%%%%%%%%%%%%%%%%%%%%%%%%%%%%%%%%%%%%%%%%
%SF refs
%%%%%%%%%%%%%%%%%%%%%%%%%%%%%%%%%%%%%%%%%%%%%%%%%%%%%%%%%

\bibitem{Luscher:1992an}
  M.~L\"uscher, R.~Narayanan, P.~Weisz and U.~Wolff,
  %``The Schrodinger functional: A Renormalizable probe for nonAbelian gauge
  %theories,''
  Nucl.\ Phys.\  B {\bf 384}, 168 (1992)
  [arXiv:hep-lat/9207009].
  %%CITATION = NUPHA,B384,168;%%

\bibitem{Luscher:1993gh}
  M.~L\"uscher, R.~Sommer, P.~Weisz and U.~Wolff,
  ``A precise determination of the running coupling in the SU(3) Yang-Mills theory,''
  Nucl.\ Phys.\  B {\bf 413}, 481 (1994)
  [arXiv:hep-lat/9309005].
  %%CITATION = NUPHA,B413,481;%%

\bibitem{Sint:1995ch}
  S.~Sint and R.~Sommer,
  ``The running coupling from the QCD Schr\"odinger functional: A one loop analysis,''
  Nucl.\ Phys.\  B {\bf 465}, 71 (1996)
  [arXiv:hep-lat/9508012].
  %%CITATION = NUPHA,B465,71;%%

\bibitem{Jansen:1998mx}
  K.~Jansen and R.~Sommer  [ALPHA collaboration],
  ``O($\alpha$) improvement of lattice QCD with two flavors of Wilson quarks,''
  Nucl.\ Phys.\  B {\bf 530}, 185 (1998)
  [Erratum-{\em ibid.}\  B {\bf 643}, 517 (2002)]
  [arXiv:hep-lat/9803017].
  %%CITATION = NUPHA,B530,185;%%

\bibitem{DellaMorte:2004bc}
  M.~Della Morte {\em et al.} [ALPHA Collaboration],
  ``Computation of the strong coupling in QCD with two dynamical flavours,''
  Nucl.\ Phys.\  B {\bf 713}, 378 (2005)
  [arXiv:hep-lat/0411025].
  %%CITATION = NUPHA,B713,378;%%

%\cite{Sint:1998iq}
\bibitem{Sint:1998iq}
  S.~Sint and P.~Weisz  [ALPHA collaboration],
  ``The running quark mass in the SF scheme and its two-loop anomalous dimension,''
  Nucl.\ Phys.\  B {\bf 545}, 529 (1999)
  [arXiv:hep-lat/9808013].
  %%CITATION = NUPHA,B545,529;%%

%\cite{Capitani:1998mq}
\bibitem{Capitani:1998mq}
  S.~Capitani, M.~L\"uscher, R.~Sommer and H.~Wittig  [ALPHA Collaboration],
  ``Non-perturbative quark mass renormalization in quenched lattice QCD,''
  Nucl.\ Phys.\  B {\bf 544}, 669 (1999)
  [arXiv:hep-lat/9810063].
  %%CITATION = NUPHA,B544,669;%%

%\cite{DellaMorte:2005kg}
\bibitem{DellaMorte:2005kg}
  M.~Della Morte {\em et al.} %R.~Hoffmann, F.~Knechtli, J.~Rolf, R.~Sommer, I.~Wetzorke and U.~Wolff
  [ALPHA Collaboration],
  ``Non-perturbative quark mass renormalization in two-flavor QCD,''
  Nucl.\ Phys.\  B {\bf 729}, 117 (2005)
  [arXiv:hep-lat/0507035].
  %%CITATION = NUPHA,B729,117;%%

%\cite{Bursa:2009we}
\bibitem{Bursa:2009we}
  F.~Bursa, L.~Del Debbio, L.~Keegan, C.~Pica and T.~Pickup,
  ``Mass anomalous dimension in SU(2) with two adjoint fermions,''
  Phys.\ Rev.\  D {\bf 81}, 014505 (2010)
  [arXiv:0910.4535 [hep-ph]].
  %%CITATION = PHRVA,D81,014505;%%
  
%%%%%%%%%%%%%%%%%%%%%%%%%%%%%%%%%%%%%%%%%%%%%%%%%%%%%%%%%
%strong coupling  refs
%%%%%%%%%%%%%%%%%%%%%%%%%%%%%%%%%%%%%%%%%%%%%%%%%%%%%%%%%
%\cite{Cardy:1996xt}
\bibitem{Cardy:1996xt}
  J.~L.~Cardy,
  {\it Scaling and Renormalization in Statistical Physics,}
(Cambridge, UK: Univ. Pr., 1996), 238 p. 

%\cite{Kennedy:1984dk}
\bibitem{Kennedy:1984dk}
  A.~D.~Kennedy, J.~Kuti, S.~Meyer and B.~J.~Pendleton,
  ``Where is the continuum in lattice quantum chromodynamics?,''
  Phys.\ Rev.\ Lett.\  {\bf 54}, 87 (1985).
  %%CITATION = PRLTA,54,87;%%

%\cite{Gottlieb:1985ug}
\bibitem{Gottlieb:1985ug}
  S.~A.~Gottlieb {\em et al.}, %J.~Kuti, D.~Toussaint, A.~D.~Kennedy, S.~Meyer, B.~J.~Pendleton and R.~L.~Sugar,
  ``The deconfining phase transition and the continuum limit of lattice quantum chromodynamics,''
  Phys.\ Rev.\ Lett.\  {\bf 55}, 1958 (1985).
  %%CITATION = PRLTA,55,1958;%%

%\cite{Farchioni:2004us}
\bibitem{Farchioni:2004us}
  F.~Farchioni {\it et al.},
  ``Twisted mass quarks and the phase structure of lattice QCD,''
  Eur.\ Phys.\ J.\  C {\bf 39}, 421 (2005)
  [arXiv:hep-lat/0406039].
  %%CITATION = EPHJA,C39,421;%%

%\cite{Catterall:2007yx}
\bibitem{Catterall:2007yx}
  S.~Catterall and F.~Sannino,
  ``Minimal walking on the lattice,''
  Phys.\ Rev.\  D {\bf 76}, 034504 (2007)
  [arXiv:0705.1664 [hep-lat]].
  %%CITATION = PHRVA,D76,034504;%%

%\cite{Catterall:2008qk}
\bibitem{Catterall:2008qk}
  S.~Catterall, J.~Giedt, F.~Sannino and J.~Schneible,
  ``Phase diagram of SU(2) with 2 flavors of dynamical adjoint quarks,''
  JHEP {\bf 0811}, 009 (2008)
  [arXiv:0807.0792 [hep-lat]].
  %%CITATION = JHEPA,0811,009;%%

%\cite{Hietanen:2008mr}
\bibitem{Hietanen:2008mr}
  A.~J.~Hietanen, J.~Rantaharju, K.~Rummukainen and K.~Tuominen,
  ``Spectrum of SU(2) lattice gauge theory with two adjoint Dirac flavours,''
  JHEP {\bf 0905}, 025 (2009)
  [arXiv:0812.1467 [hep-lat]].
  %%CITATION = JHEPA,0905,025;%%

%\cite{Hietanen:2009az}
\bibitem{Hietanen:2009az}
  A.~J.~Hietanen, K.~Rummukainen and K.~Tuominen,
  ``Evolution of the coupling constant in SU(2) lattice gauge theory with two adjoint fermions,''
  Phys.\ Rev.\  D {\bf 80}, 094504 (2009)
  [arXiv:0904.0864 [hep-lat]].
  %%CITATION = PHRVA,D80,094504;%%

%\cite{Catterall:2009sb}
\bibitem{Catterall:2009sb}
  S.~Catterall, J.~Giedt, F.~Sannino and J.~Schneible,
  ``Probes of nearly conformal behavior in lattice simulations of minimal walking technicolor,''
  arXiv:0910.4387 [hep-lat].
  %%CITATION = ARXIV:0910.4387;%%

%\cite{DelDebbio:2010hu}
\bibitem{DelDebbio:2010hu}
  L.~Del Debbio, B.~Lucini, A.~Patella, C.~Pica and A.~Rago,
  ``Mesonic spectroscopy of Minimal Walking Technicolor,''
  arXiv:1004.3197 [hep-lat].
  %%CITATION = ARXIV:1004.3197;%%

%\cite{DelDebbio:2010hx}
\bibitem{DelDebbio:2010hx}
  L.~Del Debbio, B.~Lucini, A.~Patella, C.~Pica and A.~Rago,
  ``The infrared dynamics of Minimal Walking Technicolor,''
  arXiv:1004.3206 [hep-lat].
  %%CITATION = ARXIV:1004.3206;%%

%\cite{Iwasaki:1991mr}
\bibitem{Iwasaki:1991mr}
  Y.~Iwasaki, K.~Kanaya, S.~Sakai and T.~Yoshi\'e,
  ``Quark confinement and number of flavors in strong coupling lattice QCD,''
  Phys.\ Rev.\ Lett.\  {\bf 69}, 21 (1992).
  %%CITATION = PRLTA,69,21;%%

%\cite{Nagai:2009ip}
\bibitem{Nagai:2009ip}
  K.~Nagai, G.~Carrillo-Ruiz, G.~Koleva and R.~Lewis,
  ``Exploration of SU($N_c$) gauge theory with many Wilson fermions at strong coupling,''
  Phys.\ Rev.\  D {\bf 80}, 074508 (2009)
  [arXiv:0908.0166 [hep-lat]].
  %%CITATION = PHRVA,D80,074508;%%

%\cite{Yamada:2010wd}
\bibitem{Yamada:2010wd}
  N.~Yamada {\em et al.}, %M.~Hayakawa, K.~I.~Ishikawa, Y.~Osaki, S.~Takeda and S.~Uno,
  ``Study of the running coupling constant in 10-flavor QCD with the Schr\"odinger functional method,''
  arXiv:1003.3288 [hep-lat];
  %%CITATION = ARXIV:1003.3288;%%


%%%%%%%%%%%%%%%%%%%%%%%%%%%%%%%%%%%%%%%%%%%%%%%%%%%%%%%%%
%conclusion  refs
%%%%%%%%%%%%%%%%%%%%%%%%%%%%%%%%%%%%%%%%%%%%%%%%%%%%%%%%%
%\cite{Golterman:2008gt}
\bibitem{Golterman:2008gt}
  M.~Golterman,
  %``QCD with rooted staggered fermions,''
  PoS {\bf CONFINEMENT8}, 014 (2008)
  [arXiv:0812.3110 [hep-ph]].
  %%CITATION = POSCI,CONFINEMENT8,014;%%


%%%%%%%%%%%%%%%%%%%%%%%%%%%%%%%%%%%%%%%%%%%%%%%%%%%%%%%%%
%code  refs
%%%%%%%%%%%%%%%%%%%%%%%%%%%%%%%%%%%%%%%%%%%%%%%%%%%%%%%%%
\bibitem{MILC} {\tt http://www.physics.utah.edu/$\sim$detar/milc/}

\bibitem{Hasenfratz:2008ce}
  A.~Hasenfratz, R.~Hoffmann and S.~Schaefer,
  ``Low energy chiral constants from epsilon-regime simulations with improved Wilson fermions,''
  Phys.\ Rev.\  D {\bf 78}, 054511 (2008)
  [arXiv:0806.4586 [hep-lat]].
  %%CITATION = PHRVA,D78,054511;%%

\end{thebibliography}
\end{document}